\documentclass[a4paper,fleqn]{cas-sc}

\usepackage[numbers]{natbib}

\def\tsc#1{\csdef{#1}{\textsc{\lowercase{#1}}\xspace}}
\tsc{WGM}
\tsc{QE}
\tsc{EP}
\tsc{PMS}
\tsc{BEC}
\tsc{DE}
\usepackage{graphicx}
\usepackage{amsthm}
 \usepackage{booktabs}
 \usepackage{siunitx}
 \usepackage{float}
 \usepackage{multirow}
 \usepackage{caption}
 \usepackage[table]{xcolor}
  
\begin{document}
\let\WriteBookmarks\relax
\def\floatpagepagefraction{1}
\def\textpagefraction{.001}

\shorttitle{PINNs for velocity field}

\shortauthors{Z.Tao et~al.}

\title{LSTM-PINN: An Hybrid Method for Prediction of Steady‑State Electrohydrodynamic Flow}                      
\tnotemark[1,2]


%
\author[1]{Ze Tao}[orcid=0009-0004-0202-3641]
\credit{Calculation, data analyzing, manuscript writing, review and editing}
\fnref{co-first}
\author[1]{Ke Xu}[orcid=0009-0003-7880-0235]
\credit{Calculation, data analyzing and manuscript writing}
\fnref{co-first}
\affiliation[1]{organization={Nanophotonics and Biophotonics Key Laboratory of Jilin Province, School of Physics, Changchun University of Science and Technology},
                city={Changchun},
                postcode={130022},
                country={P.R. China}}
\author[1]{Fujun Liu }[orcid=0000-0002-8573-450X]
\corref{cor}
\credit{Review and Editing}
\ead{fjliu@cust.edu.cn}
\fntext[co-first]{The authors contribute equally to this work.}
\cortext[cor]{Corresponding author}
\begin{abstract}
Physics-Informed Neural Networks (PINNs) have demonstrated considerable success in solving complex fluid dynamics problems. However, their performance often deteriorates in regimes characterized by steep gradients, intricate boundary conditions, and stringent physical constraints, leading to convergence failures and numerical instabilities. To overcome these limitations, we propose a hybrid framework that integrates Long Short-Term Memory (LSTM) networks into the PINN architecture, enhancing its ability to capture spatial correlations in the steady-state velocity field of a two-dimensional charged fluid under an external electric field. Our results demonstrate that the LSTM-enhanced PINN model significantly outperforms conventional Multilayer Perceptron (MLP)-based PINNs in terms of convergence rate, numerical stability, and predictive accuracy. This innovative approach offers improved computational efficiency and reliability for modeling electrohydrodynamic flows, providing new insights and strategies for applications in microfluidics and nanofluidics.
\end{abstract}


\begin{highlights}
\item The integration of LSTM into PINNs enables a stable computation of charged fluid velocity fields.
\item The LSTM-PINN architecture couples the Poisson-Nernst-Planck and Navier-Stokes equations.
\item Gating mechanisms within the model stabilize gradients and enhance long-range spatial transmission.
\item Clear visualization of 2D electrohydrodynamic velocity fields is achieved through LSTM-PINN.
\end{highlights}


\maketitle

\section{Introduction} 
The accurate prediction of steady-state electrohydrodynamic (EHD) flow is critical for diverse applications, ranging from microfluidic chip design~\cite{ref1} and electroosmotic micropumps~\cite{ref6} to energy harvesting~\cite{ref8} and digital single-cell analysis~\cite{ref10}. Governed by the electric double layer (EDL), EHD flows are described by coupled Poisson--Boltzmann and Navier--Stokes equations~\cite{ref11,ref12}. However, challenges such as surface charge heterogeneities and fluid property variations often induce secondary vortices, asymmetries, and instabilities~\cite{ref13}, complicating their numerical resolution. Traditional methods like finite element and spectral element techniques~\cite{ref14} rely on dense meshes to resolve EDL structures, leading to prohibitive computational costs—especially in complex geometries. While alternative approaches, such as Lagrangian techniques~\cite{ref15} and multi-ion models~\cite{ref16}, address some limitations, they remain prone to coupling inaccuracies and sensitivity to boundary conditions. Experimental methods~\cite{ref17,ref18,ref19} provide empirical insights but are difficult to reconcile with simulations. Consequently, there is a growing need for mesh-free~\cite{ref20,ref21} or surrogate modeling strategies~\cite{ref22} that balance computational efficiency with physical fidelity.  

Recent advances in deep learning offer promising solutions for such challenges, leveraging its strengths in nonlinear mapping~\cite{ref23,ref24,ref25,ref26} and hierarchical feature extraction~\cite{ref27,ref28,ref29}. Techniques like normalization and regularization have enabled stable training of large-scale networks~\cite{ref30,ref31,ref32}, while architectures such as CNNs and ResNets excel at capturing multiscale spatial features. Deep learning has demonstrated remarkable success in solving PDEs, from the Fokker--Planck equation~\cite{ref35} to nonlinear elasticity~\cite{ref36}, and has been applied to diverse domains, including hydrology~\cite{ref37} and dynamical system identification~\cite{ref38}. However, purely data-driven methods often suffer from poor physical consistency and generalization, especially when training data are scarce. This limitation underscores the necessity of hybrid approaches that integrate physical laws with data-driven learning.  

Physics-Informed Neural Networks (PINNs) address this gap by embedding governing equations directly into the neural network's loss function, enabling data-efficient and physics-consistent solutions~\cite{mcclenny2023self}. PINNs have achieved notable success across disciplines, including heat transfer~\cite{tao2025analytical}, fluid mechanics~\cite{cai2021physics}, and geophysics~\cite{liu2023joint}. Extensions like gPINNs~\cite{rasht2022physics} and $\Delta$-PINNs further enhance their capability to handle multiphysics problems~\cite{zhang2024physics}. Despite these advances, conventional PINNs based on multi-layer perceptrons (MLPs) struggle with steep gradients and complex boundary conditions, often leading to convergence failures. Long Short-Term Memory (LSTM) networks, renowned for capturing sequential dependencies~\cite{habiba2022ordinary}, have shown superior performance in dynamical systems when integrated into PINNs~\cite{liu2023novel}. However, their potential for steady-state problems remains underexplored.  

In this work, we bridge this gap by introducing pseudo-sequential representations to adapt LSTM networks for steady-state EHD flows. By reformulating spatial dependencies as learnable sequences, our method leverages LSTM's inherent strength in long-range correlation modeling—even in the absence of explicit temporal dynamics. We evaluate the proposed LSTM-PINN framework on benchmark nanofluidic transport problems, demonstrating significant improvements in predictive accuracy, convergence rate, and computational efficiency compared to traditional MLP-based PINNs. Our results establish a robust foundation for next-generation simulations in micro- and nanofluidics, with direct implications for device design and optimization.

\section{Problem Formulation}

\subsection{Governing Equations and Physical Assumptions}
We investigate the steady-state electrokinetic flow of an incompressible Newtonian electrolyte through a two-dimensional micro/nanofluidic channel cross-section. This canonical configuration captures essential electrohydrodynamic phenomena relevant to diverse applications, including electroosmotic pumping, charged nanoparticle transport, and precision microfluidic control. The formulation applies equally to nanochannel electrokinetics, nanopore ionic transport, and electroosmotic pump systems.

The computational domain is defined as the unit square $\Omega = [0, 1] \times [0, 1]$, representing a normalized channel cross-section subjected to a spatially uniform external electric field. The flow is governed by the steady incompressible Navier-Stokes equations augmented by an electrostatic body force, coupling hydrodynamics with electrostatics under local thermodynamic equilibrium.

Following Marini \emph{et al.}~\cite{marini2012charge}, the momentum balance for the charged fluid is given by:
\begin{equation}
\label{eq:full_momentum}
\partial_t \mathbf{u} + (\mathbf{u}\cdot\nabla)\mathbf{u}
= -\frac{1}{\rho_m}\nabla P
+ \frac{\rho_e}{\rho_m}\mathbf{E}
+ \frac{\eta}{\rho_m}\nabla^2\mathbf{u}
+ \frac{\tfrac{1}{3}\eta + \eta_b}{\rho_m}\nabla(\nabla\!\cdot\mathbf{u}),
\end{equation}
where $\mathbf{u}$ is the velocity field, $P$ the pressure, $\rho_m$ the mass density, $\rho_e$ the free charge density, $\mathbf{E}$ the applied electric field, and $\eta$ and $\eta_b$ the dynamic and bulk viscosities, respectively. The terms $\nabla^2 \mathbf{u}$ and $\nabla(\nabla \cdot \mathbf{u})$ represent viscous and volumetric stresses.

\subsection{Dimensional Reduction and Simplifications}
Under steady two-dimensional flow ($\mathbf{u} = (u(x,y), v(x,y))$) with constant physical parameters and negligible bulk viscosity ($\eta_b \approx 0$), the system reduces to:
\begin{align}
\label{eq:momentum_x}
\rho_m(u\partial_x u + v\partial_y u) &= \eta(\partial_{xx}u + \partial_{yy}u) 
+ \tfrac{\eta}{3}(\partial_{xy}u + \partial_{xy}v) 
+ \rho_e E_x, \\
\label{eq:momentum_y}
\rho_m(u\partial_x v + v\partial_y v) &= \eta(\partial_{xx}v + \partial_{yy}v) 
+ \tfrac{\eta}{3}(\partial_{xy}u + \partial_y v) 
+ \rho_e E_y, \\
\label{eq:continuity}
\partial_x u + \partial_y v &= 0.
\end{align}

The $x$-momentum equation (Eq.~\eqref{eq:momentum_x}) retains mixed derivatives ($\partial_{xy}v$) capturing shear-induced stresses, while Eq.~\eqref{eq:momentum_y} includes normal stress contributions ($-\tfrac{\eta}{3}(\partial_{xy}u + \partial_y v)$).

\subsection{Nondimensionalization and Boundary Conditions}
Characteristic scales normalize the system:
\begin{equation}
\label{eq:nondim}
\rho_m = 1, \quad \eta = 1, \quad \rho_e = 1, \quad E_x = 1, \quad E_y = 1,
\end{equation}
with homogeneous Neumann boundary conditions on $\partial\Omega$:
\begin{equation}
\label{eq:bc}
\frac{\partial u}{\partial n} = 0, \quad \frac{\partial v}{\partial n} = 0 \quad \text{on } \partial\Omega.
\end{equation}

\subsection{PINN Implementation Framework}
The steady-state solution is obtained by minimizing the residuals of Eqs.~\eqref{eq:momentum_x}--\eqref{eq:continuity} using a Physics-Informed Neural Network (PINN). The architecture enforces boundary conditions (Eq.~\eqref{eq:bc}) via automatic differentiation, with two variants compared:

\begin{figure}[!ht]
    \centering
    \includegraphics[width=0.8\textwidth]{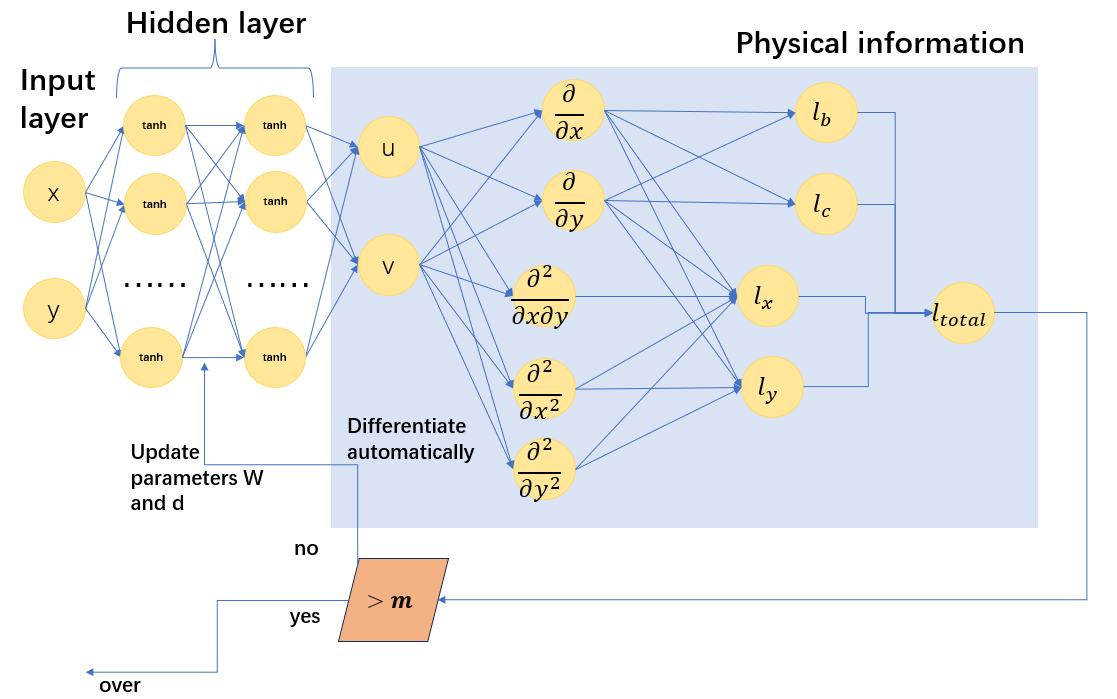}
    \caption{Schematic of the baseline MLP-based PINN architecture: the network takes spatial coordinates $(x,y)$ as input and outputs velocity components $(u,v)$. Within the PINN, we compute the residuals of the steady electrohydrodynamic momentum equations and the incompressibility constraint (Eqs.~\eqref{eq:momentum_x}–\eqref{eq:continuity}) via automatic differentiation and assemble a composite loss $\mathcal{L}_{\text{total}}=\mathcal{L}_x+\mathcal{L}_y+\mathcal{L}_c+\mathcal{L}_b$, where the boundary term $\mathcal{L}_b=\mathcal{L}_l+\mathcal{L}_r+\mathcal{L}_u+\mathcal{L}_d$ enforces homogeneous Neumann conditions (Eq.~\eqref{eq:bc}) by penalizing normal velocity gradients on the four sides. The figure highlights the baseline fully connected MLP's feed-forward mapping from coordinates to velocity; training optimizes the learnable parameters $\theta=[W,b]$ with stochastic gradient descent (SGD) over $m$ iterations, serving as a reference for Figure~\ref{fig:lstm} where an LSTM-PINN replaces the MLP to enhance modeling of spatial correlations.}
    \label{fig:mlp}
\end{figure}

\begin{figure}[!ht]
    \centering
    \includegraphics[width=0.8\linewidth]{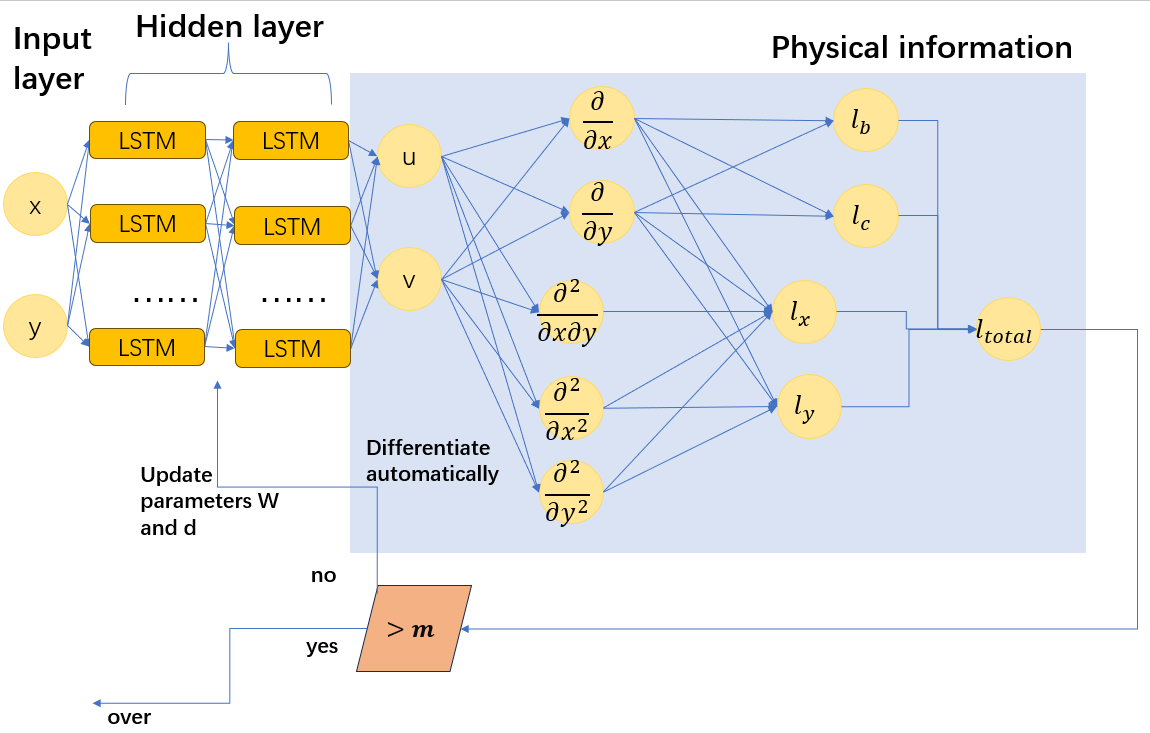}
    \caption{LSTM-PINN architecture: we replace the baseline MLP backbone with stacked LSTM layers and construct pseudo-sequential inputs from spatial coordinates so that the hidden state propagates along space and captures long-range spatial correlations. The network still maps $(x,y)$ to $(u,v)$; during training, we evaluate the residuals of the steady electrohydrodynamic momentum equations and the incompressibility constraint (Eqs.~\eqref{eq:momentum_x}–\eqref{eq:continuity})) via automatic differentiation and assemble the composite loss $\mathcal{L}_{\text{total}}=\mathcal{L}_x+\mathcal{L}_y+\mathcal{L}_c+\mathcal{L}_b$ (Eq.~\eqref{eq:total_loss}), where $\mathcal{L}_b$ enforces homogeneous Neumann boundary conditions (Eq.~\eqref{eq:bc}). Compared with Figure~\ref{fig:mlp}, this figure highlights how pseudo-sequential spatial encoding enables the LSTM's gating and memory to enhance modeling of complex spatial patterns and improve training stability for steady-state EHD flows.}
    \label{fig:lstm}
\end{figure}

\section{PINN-Based Solutions for Two-Dimensional Steady-State Electrohydrodynamic Flow Using MLP and LSTM Networks}

We employ a Physics-Informed Neural Network (PINN) framework to approximate the steady-state solution of the two-dimensional electrohydrodynamic (EHD) flow system. The baseline architecture utilizes a fully connected Multi-Layer Perceptron (MLP), shown in Figure~\ref{fig:mlp}, which takes spatial coordinates $(x,y)$ as input and outputs velocity components $(u,v)$. The governing PDE residuals are computed via automatic differentiation and incorporated into a composite loss function that enforces both the physical laws and boundary conditions.

\subsection*{Network Training and Optimization}
The trainable parameters $\theta = [W, b]$ are optimized using Stochastic Gradient Descent (SGD) over $m$ iterations. To enhance the model's capacity for capturing spatial dependencies, we replace the MLP with a LSTM network (Figure~\ref{fig:lstm}), constructing pseudo-sequential inputs to leverage the LSTM's hidden state propagation mechanism. This modification significantly improves the network's ability to resolve complex spatial patterns characteristic of EHD flows.

\subsection*{Pseudo-sequential spatial encoding and state propagation}
We reformulate spatial dependencies over $\Omega=[0,1]\times[0,1]$ into a learnable sequence to drive hidden-state propagation across space. Let $\{(x_i,y_j)\}_{i=1..N_x,\;j=1..N_y}$ denote a Cartesian grid. We impose a serpentine row-major traversal $\pi:\{1,\dots,N_xN_y\}\!\to\!\{(i,j)\}$: for odd $j$, set $t=(j-1)N_x+i$; for even $j$, set $t=(j-1)N_x+(N_x-i+1)$. This yields a sequence $\Xi=(\xi_t)_{t=1}^{N_xN_y}$ with $\xi_t=[x_t,y_t]^\top$ so that adjacent spatial points remain adjacent in the sequence, turning spatial coupling into a long-range dependency modeling problem for the LSTM. We segment $\Xi$ into row-length windows of size $L_s=N_x$ and pack rows into mini-batches $X\in\mathbb{R}^{B\times L_s\times 2}$; at the start of each window we reset $(h_0,c_0)=(0,0)$ and propagate states only within the row to avoid cross-row wrap-around. An input embedding $\phi(\xi_t)=\tanh(W_{\mathrm{in}}\xi_t+b_{\mathrm{in}})\in\mathbb{R}^{D}$ feeds the stacked LSTM, which updates according to Eqs.~\eqref{eq:cell_update}--\eqref{eq:output_gate}: We provide the compact matrix formulation of the LSTM update and a rigorous derivation of its memory dynamics in Section I of Supplementary Materials (SM) ; Eq. \eqref{eq:matrix_form} summarizes the matrix form. $c_t=f_t\odot c_{t-1}+i_t\odot g_t$, $h_t=o_t\odot\tanh(c_t)$. A linear head maps $h_t$ to $[u_t,v_t]^\top=G h_t+a$, producing velocity at each grid point; Eq.~\eqref{eq:matrix_form} provides the compact matrix form. During training, we evaluate the momentum and incompressibility residuals at each token’s $(x,y)$ and aggregate them with boundary terms into the total loss in Eq.~\eqref{eq:total_loss}. This pseudo-sequential spatial encoding couples neighborhood structure with LSTM gating, yielding the LSTM-PINN backbone in Figure~\ref{fig:lstm} for stable learning of long-range spatial correlations.

\subsection*{Loss Function Formulation}
The total loss function consists of four key components:
\begin{equation}
\mathcal{L}_{\text{total}} = \mathcal{L}_{\text{x}} + \mathcal{L}_{\text{y}} + \mathcal{L}_{\text{c}} + \mathcal{L}_{\text{b}},
\label{eq:total_loss}
\end{equation}

The $x$-momentum residual loss is computed over $n$ collocation points:
\begin{equation}
\mathcal{L}_{\text{x}} = \frac{1}{n}\sum_{i=1}^n \left[
\rho_m\left(u_i\partial_x u_i + v_i\partial_y u_i\right) 
- \eta\left(\partial_{xx}u_i + \partial_{yy}u_i\right)
- \frac{\eta}{3}\left(\partial_{xy}u_i + \partial_{xy}v_i\right)
+ \rho_e E_x
\right]^2
\label{eq:interior_loss_x}
\end{equation}

The $y$-momentum residual loss follows similarly:
\begin{equation}
\mathcal{L}_{\text{y}} = \frac{1}{n}\sum_{i=1}^n \left[
\rho_m\left(u_i\partial_x v_i + v_i\partial_y v_i\right)
- \eta\left(\partial_{xx}v_i + \partial_{yy}v_i\right)
- \frac{\eta}{3}\left(\partial_{xy}u_i + \partial_y v_i\right)
+ \rho_e E_y
\right]^2
\label{eq:interior_loss_y}
\end{equation}

The incompressibility constraint is enforced through:
\begin{equation}
\mathcal{L}_{\text{c}} = \frac{1}{n}\sum_{i=1}^n \left(\partial_x u_i + \partial_y v_i\right)^2
\label{eq:continuity_loss}
\end{equation}

The boundary loss $\mathcal{L}_{\text{b}}$ aggregates contributions from all domain boundaries:
\begin{equation}
\mathcal{L}_{\text{b}} = \mathcal{L}_{\text{l}} + \mathcal{L}_{\text{r}} + \mathcal{L}_{\text{u}} + \mathcal{L}_{\text{d}}
\label{eq:boundary_loss}
\end{equation}

Each component is evaluated over $n_1$ boundary points:
\begin{align}
\mathcal{L}_{\text{l}} &= \frac{1}{n_1}\sum_{i=1}^{n_1}\left[(\partial_x u_i)^2 + (\partial_x v_i)^2\right] \quad \text{(left: $x=0$)} \label{eq:left_boundary_loss} \\
\mathcal{L}_{\text{r}} &= \frac{1}{n_1}\sum_{i=1}^{n_1}\left[(\partial_x u_i)^2 + (\partial_x v_i)^2\right] \quad \text{(right: $x=1$)} \label{eq:right_boundary_loss} \\
\mathcal{L}_{\text{u}} &= \frac{1}{n_1}\sum_{i=1}^{n_1}\left[(\partial_y u_i)^2 + (\partial_y v_i)^2\right] \quad \text{(top: $y=1$)} \label{eq:top_boundary_loss} \\
\mathcal{L}_{\text{d}} &= \frac{1}{n_1}\sum_{i=1}^{n_1}\left[(\partial_y u_i)^2 + (\partial_y v_i)^2\right] \quad \text{(bottom: $y=0$)} \label{eq:bottom_boundary_loss}
\end{align}

\subsection*{Implementation Details}
The loss components $\mathcal{L}_{\text{x}}$ and $\mathcal{L}_{\text{y}}$ penalize momentum equation residuals, while $\mathcal{L}_{\text{c}}$ enforces fluid incompressibility. The boundary terms $\mathcal{L}_{\text{l}}$-$\mathcal{L}_{\text{d}}$ ensure compliance with homogeneous Neumann conditions by penalizing velocity gradients normal to each boundary. All derivatives are computed via automatic differentiation, and the system parameters ($\rho_m$, $\eta$, $\rho_e$, $E_x$, $E_y$) remain constant as defined in Eq.~\eqref{eq:nondim}.

\section*{LSTM Architecture and Memory Dynamics}
The LSTM network is a specialized recurrent neural network designed to overcome the gradient vanishing/explosion problems inherent in standard RNNs. Its key innovation is an internal memory state $c_t \in \mathbb{R}^D$ that propagates information through time via a gated linear update mechanism, coupled with an external hidden state $h_t \in \mathbb{R}^D$ that selectively outputs information (where $D$ denotes the state dimensionality).

The memory cell update at time $t$ is governed by:
\begin{align}
c_t &= f_t \odot c_{t-1} + i_t \odot \tilde{c}_t, \label{eq:cell_update} \\
h_t &= o_t \odot \tanh(c_t), \label{eq:hidden_update}
\end{align}
where $f_t, i_t, o_t \in [0,1]^D$ are gating vectors, and $\odot$ denotes the Hadamard product. The candidate memory state $\tilde{c}_t$ is computed as:
\begin{equation}
\tilde{c}_t = \tanh(W_c x_t + U_c h_{t-1} + b_c), \label{eq:candidate}
\end{equation}
with learnable parameters $W_c, U_c \in \mathbb{R}^{D \times M}$ and $b_c \in \mathbb{R}^D$.

\begin{figure}[t]
    \centering
    \includegraphics[width=0.9\linewidth]{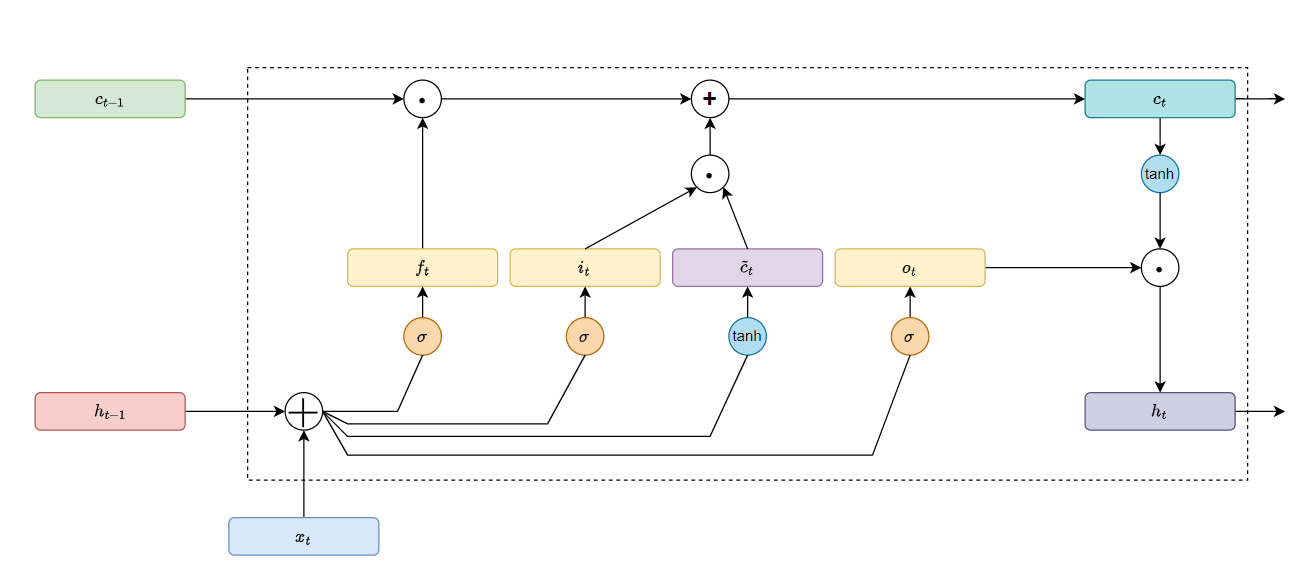}
    \caption{LSTM computational flow: the gates regulate information exchange between the cell state $c_t$ and hidden state $h_t$; the candidate memory $\tilde c_t$ arises from the current input $x_t$ and the previous hidden state $h_{t-1}$; the update obeys $c_t=f_t\odot c_{t-1}+i_t\odot \tilde c_t$ and $h_t=o_t\odot\tanh(c_t)$ (Eqs.~\eqref{eq:cell_update}--\eqref{eq:hidden_update}), with $\tilde c_t=\tanh(W_c x_t+U_c h_{t-1}+b_c)$ (Eq.~\eqref{eq:candidate}) and gate definitions $i_t=\sigma(W_i x_t+U_i h_{t-1}+b_i)$, $f_t=\sigma(W_f x_t+U_f h_{t-1}+b_f)$, $o_t=\sigma(W_o x_t+U_o h_{t-1}+b_o)$ (Eqs.~\eqref{eq:input_gate}--\eqref{eq:output_gate}). The compact matrix form in Eq.~\eqref{eq:matrix_form} consolidates parameters into $W$ and $b$. The diagram highlights how specific gate settings (e.g., $f_t\!\approx\!0, i_t\!\approx\!1$ to overwrite; $f_t\!\approx\!1, i_t\!\approx\!0$ to retain) control memory updates and, under pseudo-sequential spatial inputs, support long-range correlation modeling in the LSTM-PINN.}
    \label{fig:lstm_architecture}
\end{figure}

\subsection*{Gating Mechanism}
The LSTM employs three differentiable gates that dynamically control information flow:

\begin{align}
i_t &= \sigma(W_i x_t + U_i h_{t-1} + b_i) \quad \text{(Input gate)} \label{eq:input_gate} \\
f_t &= \sigma(W_f x_t + U_f h_{t-1} + b_f) \quad \text{(Forget gate)} \label{eq:forget_gate} \\
o_t &= \sigma(W_o x_t + U_o h_{t-1} + b_o) \quad \text{(Output gate)} \label{eq:output_gate}
\end{align}

where $\sigma$ denotes the sigmoid function. These gates enable precise control over memory operations:
\begin{itemize}
    \item When $f_t \approx 0$ and $i_t \approx 1$, the cell discards old memory and stores new information
    \item When $f_t \approx 1$ and $i_t \approx 0$, the cell preserves its current state
\end{itemize}

\subsection*{Compact Representation}
The complete LSTM update can be expressed efficiently in matrix form:
\begin{equation}
\begin{bmatrix}
\tilde{c}_t \\ o_t \\ i_t \\ f_t
\end{bmatrix}
=
\begin{bmatrix}
\tanh \\ \sigma \\ \sigma \\ \sigma
\end{bmatrix}
\left(
W 
\begin{bmatrix}
x_t \\ h_{t-1}
\end{bmatrix}
+ b
\right), \label{eq:matrix_form}
\end{equation}
where $W \in \mathbb{R}^{4D \times (M+D)}$ and $b \in \mathbb{R}^{4D}$ consolidate all trainable parameters.

\subsection*{Memory Hierarchy}
The LSTM implements a multi-scale memory system:
\begin{itemize}
    \item \textbf{Short-term}: Hidden state $h_t$ provides immediate, transient representation
    \item \textbf{Intermediate}: Cell state $c_t$ maintains adaptable, sequence-spanning context
    \item \textbf{Long-term}: Network weights encode persistent, task-specific knowledge
\end{itemize}

This hierarchy enables the LSTM to model dependencies across varying time scales - the cell state $c_t$ bridges the gap between transient activations ($h_t$) and fixed parameters, justifying the "Long Short-Term Memory" nomenclature. The architecture is particularly effective for learning spatial correlations in steady-state systems when adapted through pseudo-sequential inputs.

\section{Results and Discussions}

\subsection{Loss Convergence Analysis of LSTM and MLP Architectures Under Variable Learning Rates}

\begin{figure}[t]
    \centering
    \includegraphics[width=0.95\linewidth]{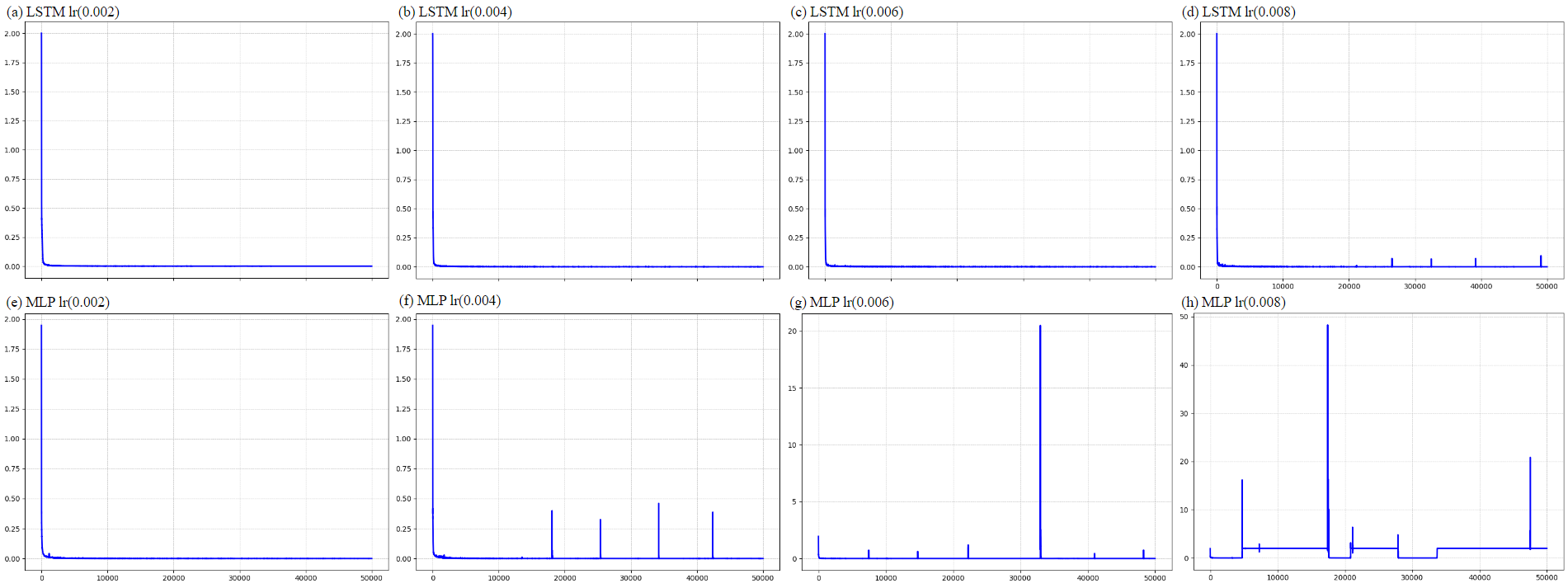}
    \caption{MLP-PINN loss trajectories at learning rates $\lambda\in\{0.002,0.004,0.006,0.008\}$, using a common axis range for side-by-side comparison. The curves reveal strong learning-rate sensitivity: oscillations intensify at $\lambda=0.004$, large-amplitude fluctuations emerge at $\lambda=0.006$, and optimization becomes chaotic at $\lambda=0.008$, which halts training.}
    \label{fig:mlp_convergence}
\end{figure}

\begin{figure}[t]
    \centering
    \includegraphics[width=0.95\linewidth]{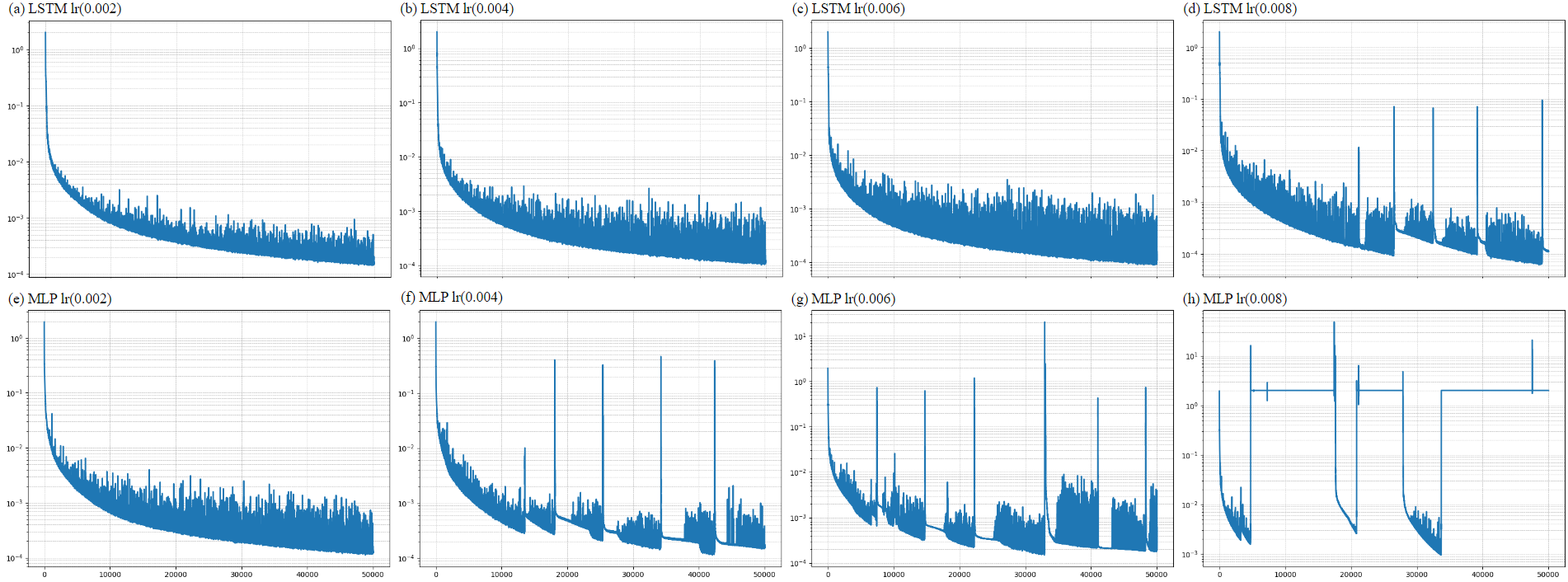}
    \caption{LSTM-PINN loss trajectories at learning rates $\lambda\in\{0.002,0.004,0.006,0.008\}$, displayed on a common axis range for side-by-side comparison. The curves exhibit robust, smooth convergence across the intermediate rates ($0.002$--$0.005$) and show only minor terminal variations at the higher rates ($0.006$--$0.008$), reflecting stable training dynamics in the LSTM-based PINN.}
    \label{fig:lstm_convergence}
\end{figure}

\begin{figure}[t]
    \centering
    \includegraphics[width=0.95\linewidth]{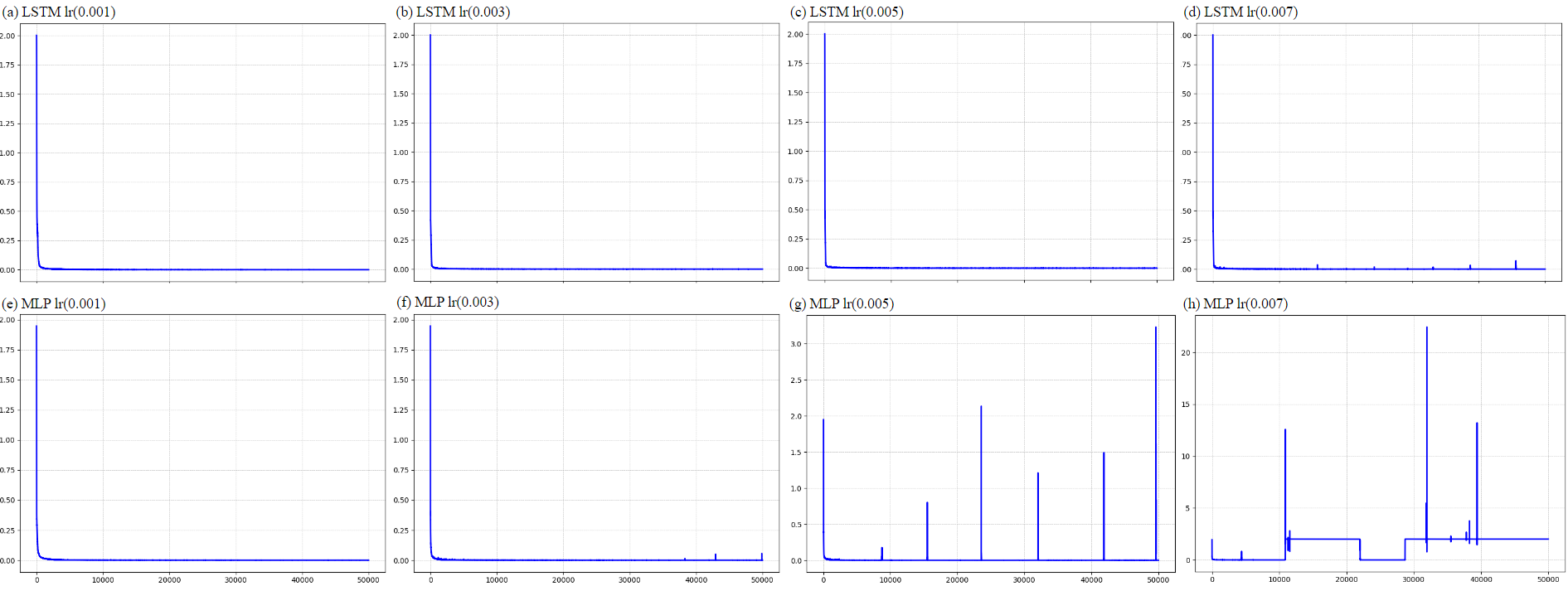}
    \caption{We compare LSTM-PINN and MLP-PINN at four learning rates $\lambda=\{0.001,0.003,0.005,0.007\}$. The LSTM-PINN descends smoothly to $10^{-3}$ at $\lambda=0.001$, sustains robust behavior over $\lambda=0.003$--$0.005$ with the fastest convergence at $\lambda=0.005$, and maintains functional training at $\lambda=0.007$ with only minor terminal variations. The MLP-PINN stabilizes only at $lambda=0.001$; low-frequency oscillations appear at $\lambda=0.003$, and the fluctuations intensify at higher rates, trending toward instability. This side-by-side panel characterizes the learning-rate tolerance and convergence traits of the two architectures within the PINN framework.}
    \label{fig:comparative}
\end{figure}

\begin{figure}[t]
    \centering
    \includegraphics[width=0.95\linewidth]{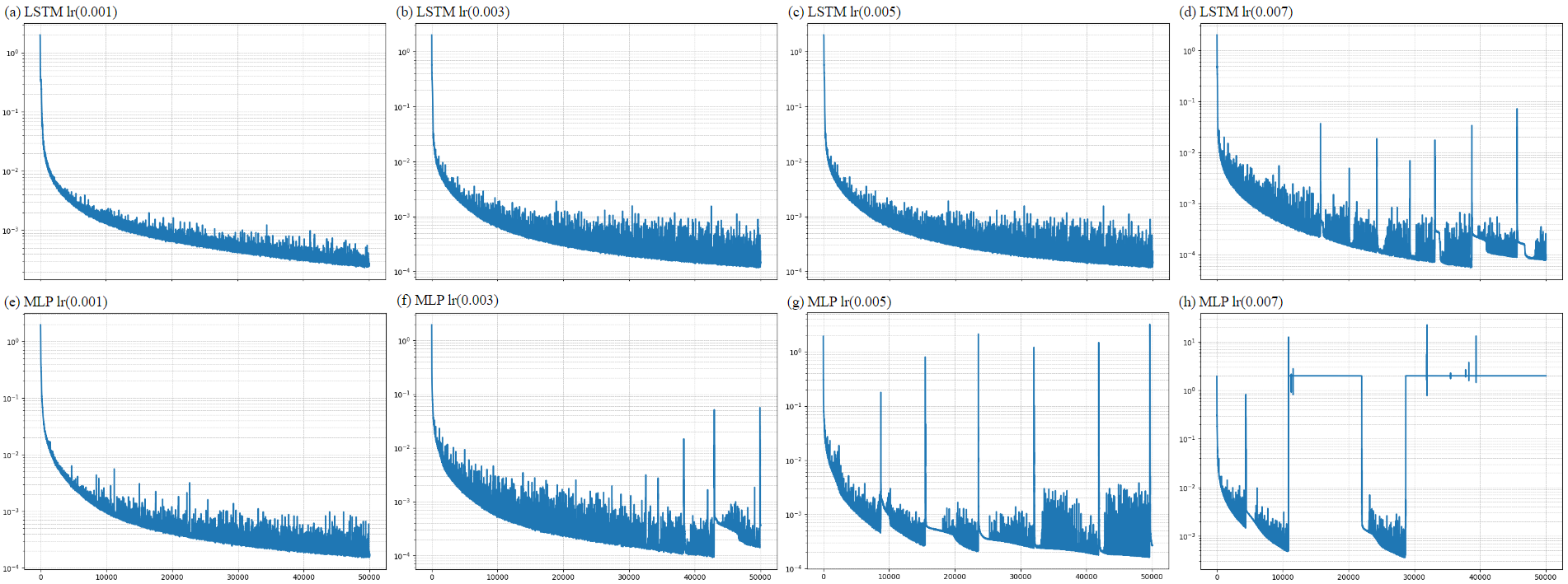}
    \caption{We train LSTM-PINN (top row) and MLP-PINN (bottom row) for $5\times10^4$ steps under identical physics, sampling, and loss composition, and we align columns by learning rate $\lambda\in\{0.001,0.003,0.005,0.007\}$: (a)(e) $\lambda=0.001$, (b)(f) $0.003$, (c)(g) $0.005$, (d)(h) $0.007$. We plot step-wise total loss on a common vertical scale. LSTM-PINN descends smoothly for $0.001$--$0.005$ and drives the baseline to $10^{-3}$--$10^{-4}$; at $\lambda=0.007$ it still trends downward with intermittent spikes and short plateaus. MLP-PINN converges slowly at $\lambda=0.001$; at $\lambda=0.003$ it shows low-frequency oscillations and late spikes; at $\lambda=0.005$ it cycles through ``fast drop--spike--reset'' phases; at $\lambda=0.007$ it exhibits large spikes and long plateaus. This same-scale panel enables a direct comparison of the two backbones' responses to the learning rate.}
    \label{c21}
\end{figure}

The convergence analysis reveals distinct training behaviors between MLP and LSTM architectures within the PINN framework. Figure~\ref{fig:mlp_convergence} demonstrates the MLP's sensitivity to learning rate selection, where stable convergence to $10^{-4}$ occurs only at $\lambda=0.001$. At $\lambda=0.003$, low-frequency oscillations emerge, indicating early gradient instability. These oscillations intensify significantly at $\lambda=0.004$, evolving into large-amplitude fluctuations in the $0.006$--$0.007$ range. Complete convergence failure occurs at $\lambda\geq0.008$, characterized by chaotic optimization behavior.

In contrast, Figure~\ref{fig:lstm_convergence} shows the LSTM-based PINN achieves smooth convergence to $10^{-3}$ at $\lambda=0.001$ and maintains robust performance across intermediate rates ($0.002$--$0.005$). The architecture demonstrates particular strength at $\lambda=0.005$, achieving optimal convergence speed without sacrificing stability. Remarkably, the LSTM maintains functional training even at $\lambda=0.007$, exhibiting only minor terminal variations, with instability becoming noticeable only at $\lambda=0.009$.

The direct comparison in Figure~\ref{fig:comparative} and Figure~\ref{c21} quantifies the LSTM's superior tolerance to aggressive learning rates. The MLP architecture begins diverging above $\lambda=0.004$, while the LSTM maintains stable convergence up to $\lambda=0.007$ - representing a 75\% increase in usable learning rate range. This enhanced stability originates from the LSTM's sophisticated gating mechanisms, which provide three critical functions: dynamic gradient regulation through forget/input gates, controlled information propagation via output gates, and effective mitigation of vanishing/exploding gradient problems.

These architectural features enable more reliable learning of physical patterns, particularly for challenging cases involving high-order derivatives and coupled differential constraints. The MLP's simpler structure, lacking such regulatory mechanisms, proves significantly more vulnerable to training instabilities, especially under aggressive optimization schedules.

The experimental results demonstrate a fundamental trade-off in PINN design between computational efficiency and training stability. The MLP architecture offers lower computational cost but suffers from limited stability margins. Conversely, the LSTM variant requires greater memory overhead but delivers superior convergence properties and robustness.

For complex physical systems like electrohydrodynamic flows, where coupled Poisson-Navier-Stokes equations demand stable gradient propagation, the LSTM's architectural advantages justify its additional computational requirements. The observed performance differences have particular significance for real-world applications where training efficiency and solution reliability are paramount considerations.
\begin{table}[t]
  \centering
  \captionsetup{labelfont={color=red}, textfont={color=red}}
  \caption{Training cost and final loss versus learning rate for MLP-PINN and LSTM-PINN. Each setting trains for 50,000 epochs; we report total wall-clock time (s), average time per epoch (s/epoch = total/50{,}000), and the final loss (last entry in each loss file).}
  \label{tab:training-time-loss-mlp-lstm}
  \begin{tabular}{ccccccc}
    \toprule
    $\lambda$ & \multicolumn{3}{c}{MLP-PINN} & \multicolumn{3}{c}{LSTM-PINN} \\
    \cmidrule(lr){2-4} \cmidrule(lr){5-7}
           & training time (s) & time/epoch (s) & final loss
           & training time (s) & time/epoch (s) & final loss \\
    \midrule
    1e-3 & 10279.548 & 0.205591 & 0.000171724 & 18902.793 & 0.378056 & 0.0002607590 \\
    2e-3 & 10271.793 & 0.205436 & 0.000128218 & 19384.613 & 0.387692 & 0.0001895340 \\
    3e-3 & 10253.194 & 0.205064 & 0.000163672 & 19368.632 & 0.387373 & 0.0001217320 \\
    4e-3 & 10282.871 & 0.205657 & 0.000156129 & 19399.317 & 0.387986 & 0.0001163000 \\
    5e-3 & 10289.191 & 0.205784 & 0.000137615 & 19386.729 & 0.387735 & 0.0001005610 \\
    6e-3 & 10281.259 & 0.205625 & 0.000160153 & 19361.415 & 0.387228 & 0.0000986568 \\
    7e-3 & 10279.818 & 0.205596 & 2.000000000 & 18965.927 & 0.379319 & 0.0001419920 \\
    8e-3 & 10306.271 & 0.206125 & 2.000000000 & 18974.817 & 0.379496 & 0.0000472510 \\
    \bottomrule
  \end{tabular}
  \label{t1}
\end{table}

In Table \ref{t1}, we quantify computational cost and final loss under identical hardware and training settings and then draw the backbone-level conclusion. The per-epoch time remains nearly constant with respect to the learning rate: MLP-PINN stays within \( 0.205\text{-}0.206 \, \text{s/epoch} \), and LSTM-PINN stays within \( 0.379\text{-}0.388 \, \text{s/epoch} \). Despite the larger per-step cost, LSTM achieves lower and more stable final losses in the mid-to-high learning-rate regime: for \( \lambda \geq 3 \times 10^{-3} \) the LSTM final loss is strictly smaller than the MLP counterpart (e.g., \( \lambda = 3 \times 10^{-3} \): \( 1.217 \times 10^{-4} \) vs \( 1.637 \times 10^{-4} \); \( \lambda = 6 \times 10^{-3} \): \( 9.866 \times 10^{-5} \) vs \( 1.602 \times 10^{-4} \); \( \lambda = 8 \times 10^{-3} \): \( 4.7251 \times 10^{-5} \) vs \( 2 \)), and LSTM continues to converge at \( \lambda = 7 \times 10^{-3} \) and \( 8 \times 10^{-3} \) whereas MLP logs \( 2 \) (divergence). At small rates (\( \lambda = 1\times10^{-3}, \, 2 \times 10^{-3} \)), MLP attains smaller final losses (\( 1.717 \times 10^{-4}, \, 1.282 \times 10^{-4} \) vs LSTM's \( 2.608 \times 10^{-4}, \, 1.895 \times 10^{-4} \)), and the crossover occurs around \( \lambda \approx 3 \times 10^{-3} \). 

We additionally evaluate a GRU-PINN and three recent PINN variants—self-adaptive PINN, gradient-enhanced PINN, and residual-attention PINN—under the same protocol, and report loss trajectories, steady fields, and computational cost; see Sections II-V in Supplementary Materials (SM).
\subsection{Model Fidelity and Predictive Symmetry in PINN-Based Velocity Field Reconstruction}

\begin{figure}[t]
    \centering
    \includegraphics[width=0.95\linewidth]{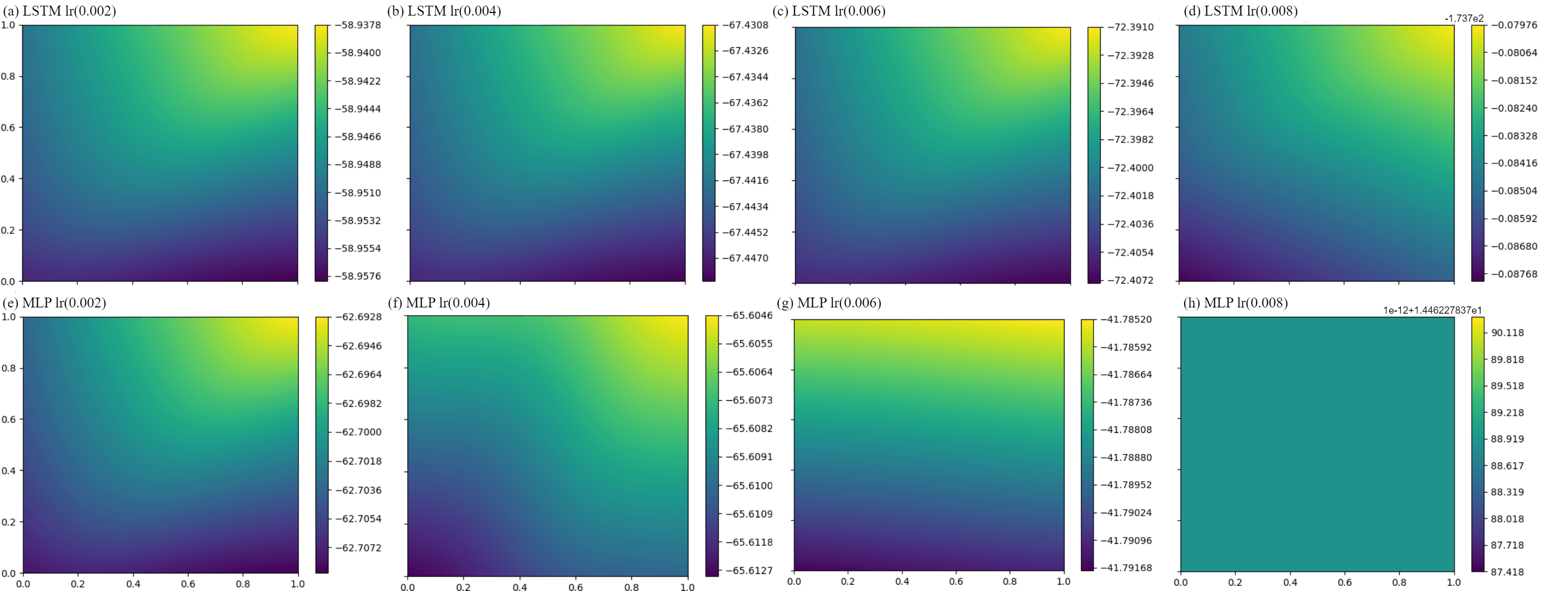}
    \caption{Predicted $x$-velocity fields at learning rates $\lambda=0.002,0.004,0.006,0.008$. We place LSTM-PINN and MLP-PINN side by side for direct inspection. The LSTM-PINN preserves the expected diagonal symmetry and smooth spatial variation across $\lambda=0.002$--$0.005$ and shows only mild magnitude modulation at $\lambda=0.006$--$0.008$; the MLP-PINN develops boundary asymmetry and localized artifacts as the rate increases, with symmetry degradation becoming pronounced over $\lambda=0.004$--$0.007$. These $x$-component panels demonstrate the LSTM's advantage in maintaining physical consistency and spatial coherence.}
    \label{fig:x_velocity}
\end{figure}
\begin{figure}[t]
    \centering
    \includegraphics[width=0.95\linewidth]{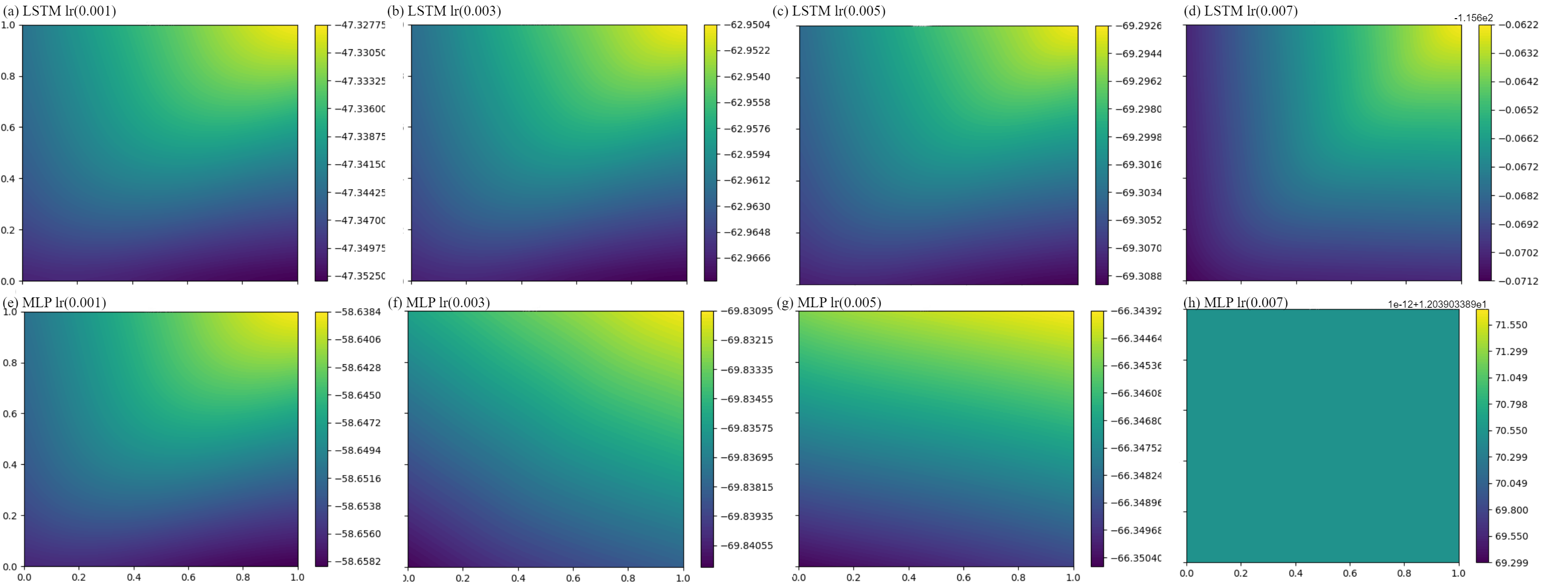}
    \caption{Predicted $x$-velocity contour maps (top row: LSTM-PINN; bottom row: MLP-PINN; columns aligned by learning rate: (a)(e) $\lambda=0.001$, (b)(f) $0.003$, (c)(g) $0.005$, (d)(h) $0.007$). We train both backbones under identical physics, sampling, and loss composition, and we plot the predicted field $u(x,y)$. The LSTM-PINN preserves a smooth, diagonally symmetric pattern over $\lambda=0.001$--$0.005$ and still retains the dominant gradient with mild magnitude modulation at $\lambda=0.007$. The MLP-PINN reproduces the baseline shape at $\lambda=0.001$; as the rate increases to $\lambda=0.003$--$0.005$ it strengthens boundary distortions and reduces global smoothness; at $\lambda=0.007$ panel (h) collapses to a near-constant field that fails to reflect the expected spatial gradient. Using a common color scale, the panel contrasts the two backbones' responses to the learning rate and demonstrates the robustness of the LSTM-PINN at the level of spatial solution quality.}
    \label{c23}
\end{figure}

\begin{figure}[t]
    \centering
    \includegraphics[width=0.95\linewidth]{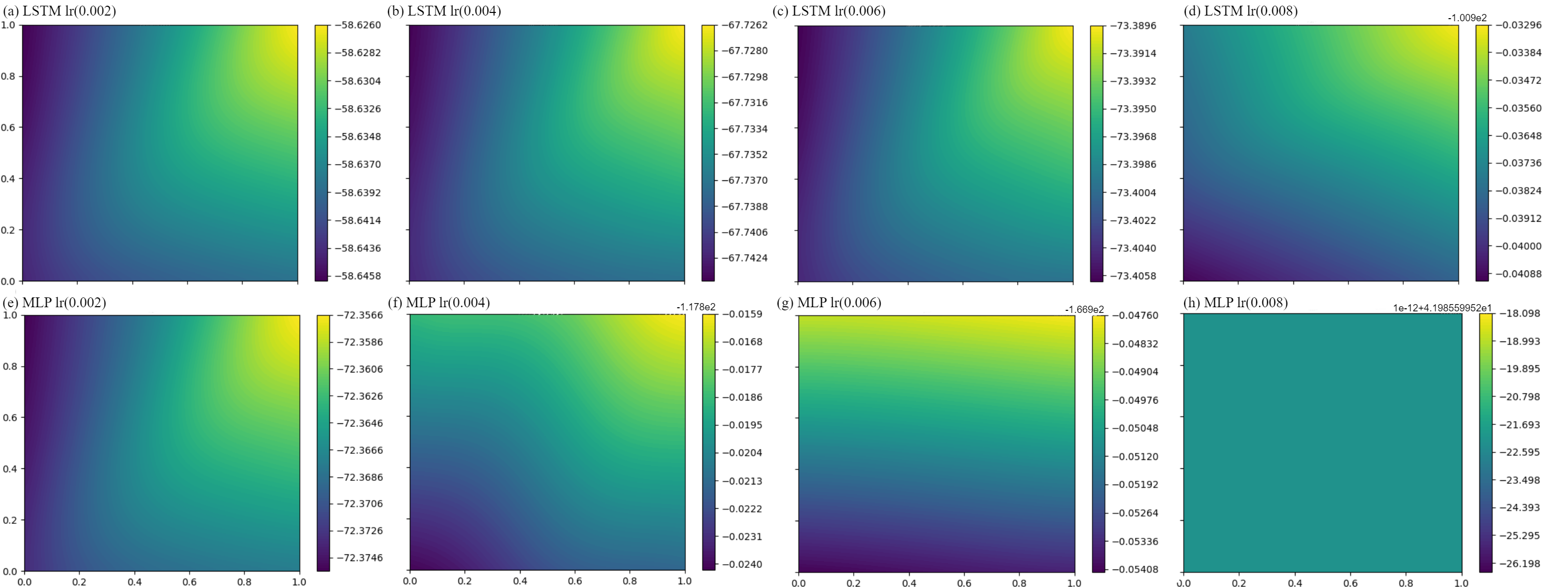}
    \caption{Predicted $y$-velocity fields at learning rates $\lambda=0.002,0.004,0.006,0.008$ (top row: LSTM-PINN; bottom row: MLP-PINN). The panels compare the two architectures under identical physics and loss settings. LSTM-PINN preserves physical consistency and spatial coherence across the sweep; MLP-PINN produces boundary asymmetry and localized artifacts as the rate increases, and the reconstruction quality degrades accordingly. These $y$-component results align with the $x$-component observations in Figures.~\ref{fig:x_velocity}-\ref{c23} and reinforce the learning-rate robustness of the LSTM design.}
    \label{fig:y_velocity}
\end{figure}
\begin{figure}[t]
    \centering
    \includegraphics[width=0.95\linewidth]{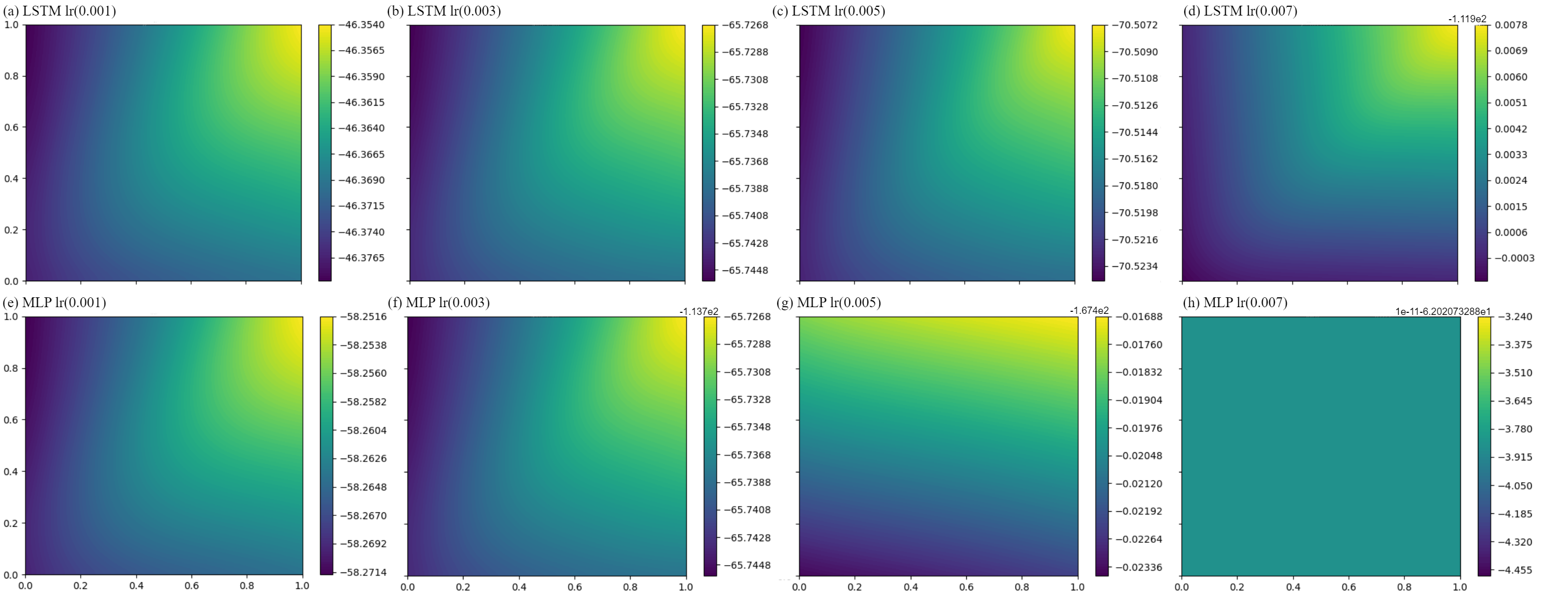}
    \caption{Predicted $y$-velocity contour maps (top row: LSTM-PINN; bottom row: MLP-PINN; columns aligned by learning rate: (a)(e) $\lambda=0.001$, (b)(f) $0.003$, (c)(g) $0.005$, (d)(h) $0.007$). We train both backbones under identical physics, sampling, and loss composition, and we plot the predicted field $v(x,y)$. The LSTM-PINN sustains a smooth, diagonally oriented dominant gradient over $\lambda=0.001$--$0.005$ and retains spatial coherence at $\lambda=0.007$ with only mild magnitude modulation. The MLP-PINN reproduces the baseline pattern at $\lambda=0.001$; as the rate increases to $\lambda=0.003$--$0.005$, boundary distortions strengthen and global smoothness weakens; at $\lambda=0.007$ in panel (h), the field collapses toward a near-constant map and fails to reflect the expected spatial gradient. This side-by-side view compares the two backbones' responses to the learning rate and, through the $y$ component, confirms the LSTM-PINN's physical consistency and robustness.}
    \label{c24}
\end{figure}
The steady-state EHD flow solution should exhibit perfect diagonal symmetry in both velocity components due to uniform boundary conditions and applied electric field. Our results in Figures.~\ref{fig:x_velocity}-\ref{c24}  demonstrate that the LSTM-based PINN maintains this essential physical symmetry across learning rates from 0.002 to 0.005, producing velocity fields with smooth spatial variations that strictly adhere to boundary conditions. While slight magnitude variations emerge at higher rates (0.006-0.008), the fundamental symmetric structure remains intact. Only at the extreme rate of 0.009 does the symmetry begin to degrade, indicating the limits of the LSTM's generalization capacity under aggressive optimization.

In stark contrast, the MLP-based PINN fails to preserve these physical constraints consistently. At moderate learning rates (0.001-0.003), while the solutions appear smooth, they already show early signs of boundary asymmetry. This physical inconsistency becomes pronounced between 0.004 and 0.007, where the velocity fields develop visible deviations from symmetric behavior and exhibit localized artifacts in high-gradient regions. The physical coherence completely breaks down at 0.008-0.009, with the velocity fields degenerating into irregular, distorted patterns that violate fundamental conservation principles.

These velocity field predictions corroborate our earlier convergence analysis, revealing a fundamental difference in how each architecture handles physics-constrained learning. The LSTM's gating mechanisms provide inherent regulation of gradient flow, enabling it to maintain spatial and physical coherence throughout the domain. The MLP, despite its computational simplicity, proves inadequate for learning coupled PDE solutions without extensive hyperparameter tuning. The consistent superiority of LSTM-based networks in both convergence behavior and solution quality establishes them as particularly suitable for multiscale, physically coupled systems where robustness and accuracy are paramount.

\vspace{1em}
\subsection{Underlying Architectural Factors Affecting Model Performance in PINN Frameworks}

The superior performance of LSTM networks over MLPs in PINN stems from fundamental architectural differences in handling spatial dependencies and enforcing physical constraints. While originally designed for sequential data processing, LSTM's memory mechanisms and recurrent structure prove equally effective at capturing long-range spatial interactions in steady-state systems. This capability becomes particularly valuable when solving PDE-constrained problems like electrohydrodynamic flows, where the network must maintain consistency across distributed collocation points. The LSTM's gating mechanisms - including forget, input, and output gates - provide dynamic control over information flow, effectively mitigating common optimization challenges such as vanishing or exploding gradients. This inherent regulation enables stable training even when dealing with the stiff gradient landscapes characteristic of high-order derivative terms in the governing equations.

In contrast, the simpler MLP architecture operates purely through pointwise transformations without any capacity for contextual memory or information regulation. This fundamental limitation manifests in several practical shortcomings: reduced ability to enforce global physical constraints, poor generalization near domain boundaries, and heightened sensitivity to hyperparameter choices. Where LSTM networks can maintain stable training across a broad range of learning rates, MLPs require careful tuning of both learning rates and network depth to avoid convergence failures or physically inconsistent solutions. The MLP's lack of internal state propagation mechanisms makes it particularly vulnerable to instability when solving coupled systems of equations, as it cannot efficiently reconcile local predictions with domain-wide conservation laws.

The comparative results demonstrate that while both architectures are theoretically capable of approximating PDE solutions, LSTM-based PINNs achieve significantly better solution fidelity and numerical robustness. This performance advantage comes from the LSTM's ability to preserve solution smoothness, maintain physical symmetries, and coherently enforce boundary conditions throughout the domain. These findings highlight the critical role of network architecture selection in physics-informed machine learning, particularly for multiscale problems involving coupled physical phenomena. For complex systems like electrohydrodynamic flows, the LSTM's enhanced stability and generalization capabilities justify its additional computational overhead, offering a favorable balance between model complexity and predictive performance.

\section{Equal-Capacity Ablation: Hidden-Width Reduction vs. Embedding/Head Reallocation—Convergence Dynamics and Baseline Suitability}
We fix the learning rate at $\lambda=5\times10^{-3}$ and keep the physics, sampling, and optimizer strictly unchanged, then design an equal-capacity ablation to isolate the marginal effects of gated memory bandwidth versus peripheral projection capacity. We reduce the hidden width of all three stacked LSTM layers from $32$ to $16$ and compensate parameters so that every variant has the same total trainables $P$. In one configuration we enlarge the input embedding, mapping $(x,y)\in\mathbb{R}^2$ to $\mathbb{R}^{298}$ via a $\tanh$ layer before feeding a $3\times$ LSTM-16 stack and a linear head to $[u,v]^\top$. In the other configuration we keep the original $2\!\to\!32$ embedding and the $3\times$ LSTM-16 backbone but append a two-layer regression head $16\!\to\!940\!\to\!2$ with an intermediate $\tanh$ so that the overall parameter count exactly matches the first configuration. We predict $(u,v)$ on $\Omega=[0,1]^2$, obtain first/second derivatives by automatic differentiation, build the momentum residuals $r_x,r_y$ and incompressibility $r_c=\partial_x u+\partial_y v$, enforce homogeneous Neumann boundaries on all four edges, and assemble the unchanged composite loss $\mathcal{L}_{\mathrm{total}}=\mathcal{L}_{x}+\mathcal{L}_{y}+\mathcal{L}_{c}+\mathcal{L}_{b}.$
We keep the interior/boundary sampling budgets fixed with per-step resampling, train for $50{,}000$ steps with Adam at $\lambda=5\times10^{-3}$, log the loss trajectory and wall-clock cost, and render the final $(u,v)$ fields on a high-resolution grid for side-by-side evaluation. Under identical physics and training schedule and with exactly matched capacity, this protocol replaces capacity in two orthogonal ways—narrowing the gated pathway while enlarging either the input embedding or the output head—so we can quantify how shrinking the hidden state affects the LSTM updates:
\begin{equation}
    c_t=f_t\odot c_{t-1}+i_t\odot g_t,\qquad h_t=o_t\odot\tanh(c_t),
\end{equation}
and how shifting parameters between the embedding and the head influences convergence speed, stability, and steady-state field structure.
\begin{figure}[t]
    \centering
    \includegraphics[width=0.95\linewidth]{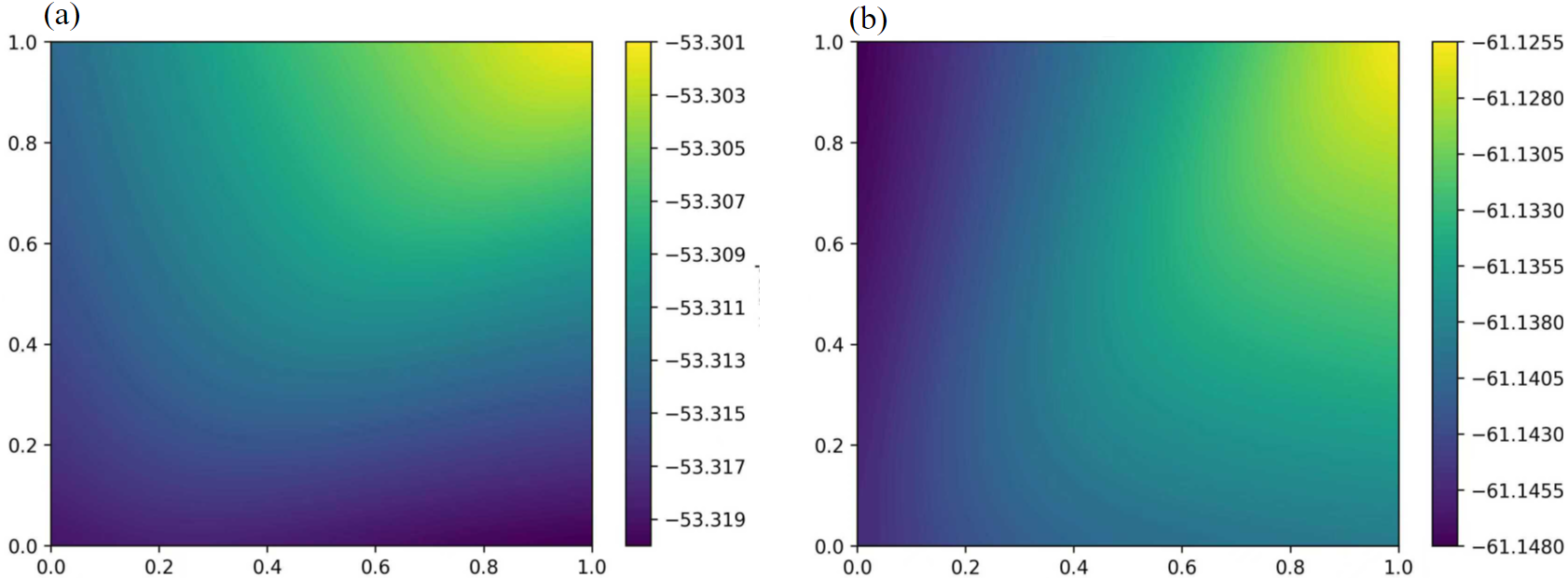}
    \caption{Under the same setting, we visualize the steady fields at step $50{,}000$: (a) $u(x,y)$; (b) $v(x,y)$. Both maps exhibit smooth, coherent gradients over $\Omega=[0,1]^2$ with no spurious boundary oscillations; the equal-parameter design preserves well-structured spatial patterns.}
    \label{YR}
\end{figure}
\begin{figure}[t]
    \centering
    \includegraphics[width=0.95\linewidth]{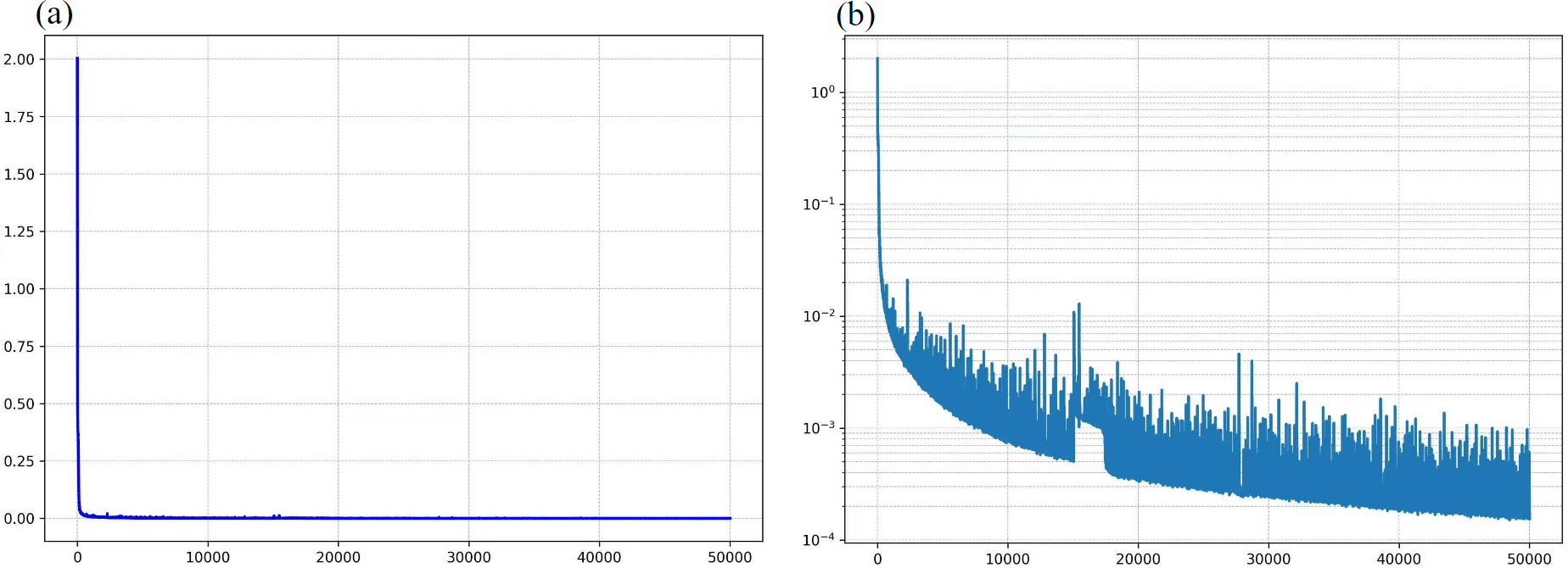}
    \caption{We reduce the hidden width of the three-layer LSTM to 16 and expand the input embedding to 298 to match the total parameter count; we train for $50{,}000$ steps at $\lambda=0.005$ and plot the total loss. (a) Linear scale—rapid drop within the first $10^3$ steps from $\mathcal{O}(1)$ to below $10^{-2}$, then persistently low; (b) Logarithmic scale—broad-range descent with a tightening noise envelope and occasional spikes, stabilizing around $10^{-4}$ toward the end.}
    \label{YL}
\end{figure}
\begin{figure}[t]
    \centering
    \includegraphics[width=0.95\linewidth]{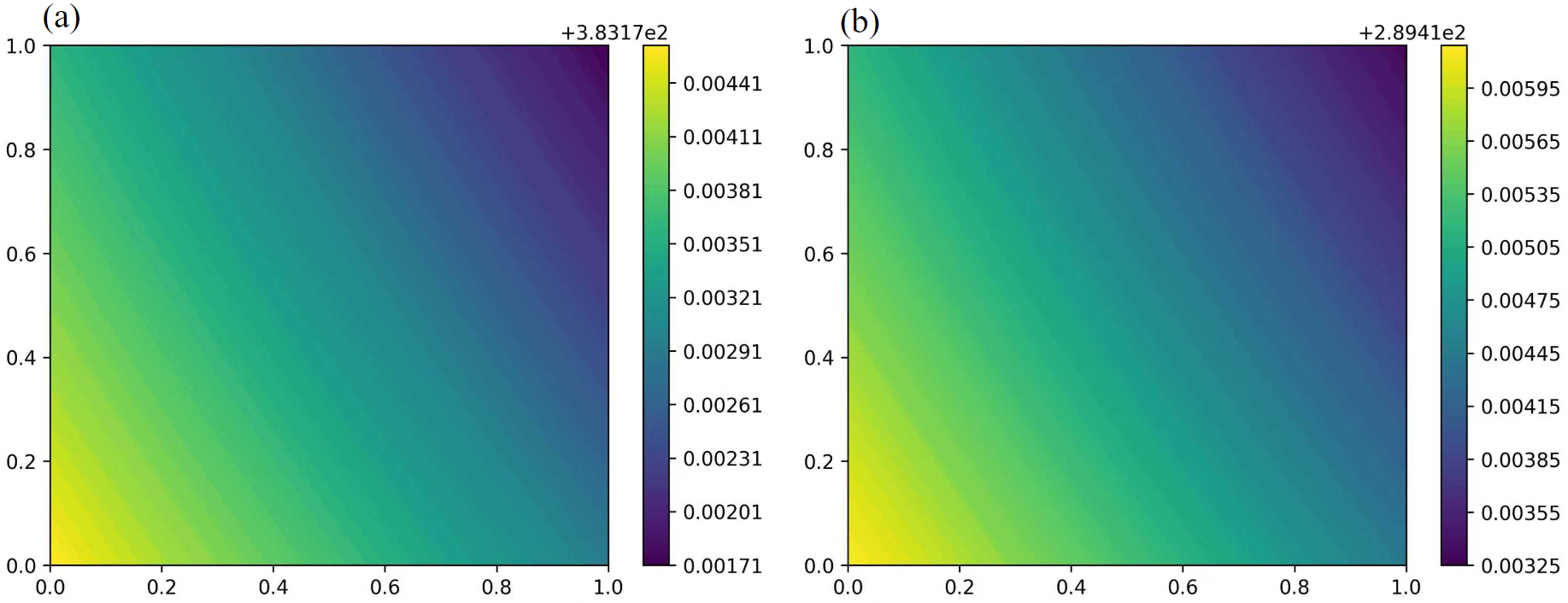}
    \caption{We keep the 32-dimensional input embedding unchanged, reduce each of the three stacked LSTM layers to hidden width 16, and attach a feed-forward head $16\!\to\!940\!\to\!2$ to match the total parameter count. At $\lambda=0.005$ for $50{,}000$ steps, we visualize the steady fields at step $50{,}000$: (a) $u(x,y)$ and (b) $v(x,y)$. Both panels show smooth, coherent gradients over $\Omega=[0,1]^2$ without spurious boundary oscillations; the equal-capacity design preserves well-structured spatial patterns.}
    \label{PR}
\end{figure}
\begin{figure}[t]
    \centering
    \includegraphics[width=0.95\linewidth]{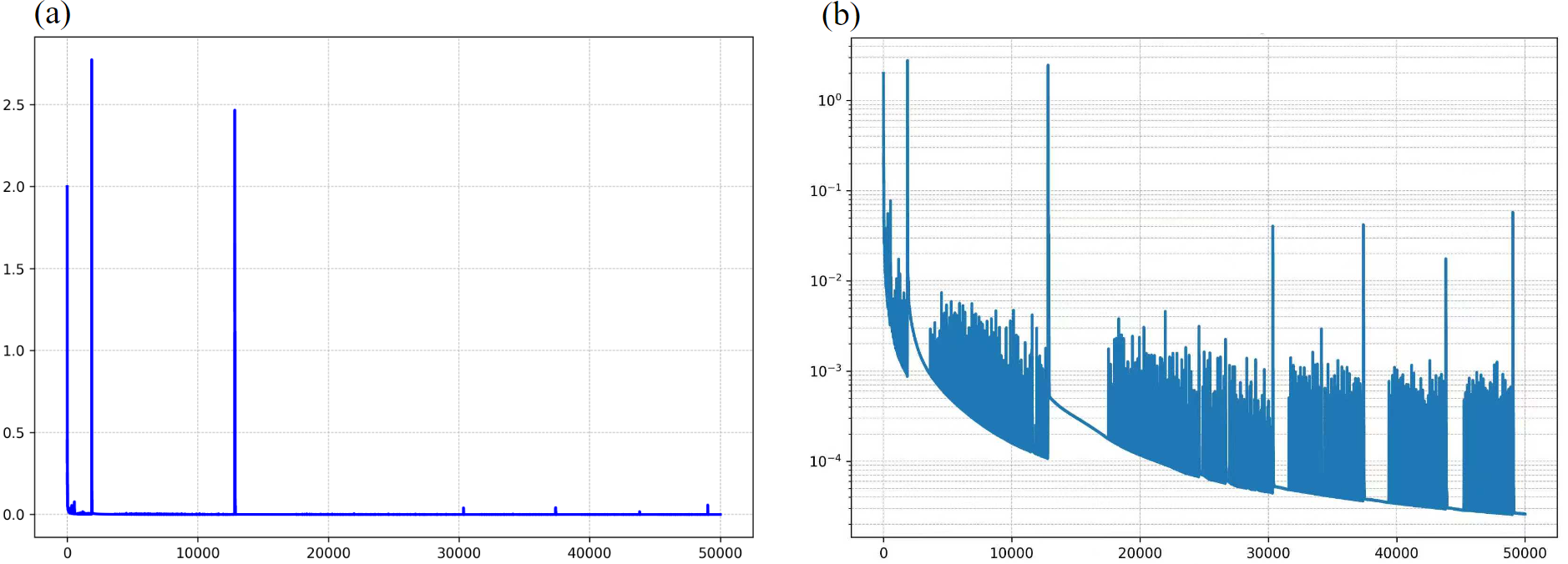}
    \caption{Under the same setting, we plot the total loss versus step: (a) linear scale and (b) logarithmic scale. After several early spikes, the trajectory enters a broad-range descent with a tightening noise envelope; occasional late spikes do not overturn the overall trend, and the curve stabilizes near $10^{-4}$ toward the end. This result shows that the “narrowed hidden state + enlarged output head” replacement maintains controlled convergence dynamics under equal capacity.}
    \label{PL}
\end{figure}
\begin{table*}[t]
  \centering
  \caption{Equal-capacity ablation of the LSTM-PINN at the fixed learning rate $\lambda=0.005$ and identical physics, sampling, optimizer, and hardware. Each model trains for $50{,}000$ epochs. We compare three configurations: the baseline LSTM-PINN; a variant that reduces all LSTM hidden widths to $16$ while enlarging the input embedding to $298$ to match the total parameter budget; and a variant that also uses hidden width $16$ but keeps the original input embedding while enlarging the output head to $16\!\to\!940\!\to\!2$ so that the total trainables remain equal. We report the wall-clock training time (s), the mean time per epoch (s/epoch $=$ total$/50{,}000$), and the final loss taken as the last item in the loss log. All numerics appear in decimal (no scientific notation) to facilitate direct, reproducible comparison.}
  \label{tab:ablation_equal_capacity_lstm_pinn}
  \begin{tabular*}{\textwidth}{@{\extracolsep{\fill}} l r r r}
    \toprule
    Variant & training time (s) & time/epoch (s) & final loss \\
    \midrule
    Hidden$=16$ & 14928.480 & 0.298570 & 0.0001570661 \\
    Hidden$=16$ & 16181.268 & 0.323625 & 0.00002591088 \\
    LSTM-PINN baseline & 19386.729 & 0.387735 & 0.0001005610 \\
    \bottomrule
  \end{tabular*}
\end{table*}

In Figures \ref{YR}-\ref{PL} and Table \ref{tab:ablation_equal_capacity_lstm_pinn}, we evaluate an equal-capacity ablation under identical physics, sampling, optimizer, and schedule ($50{,}000$ steps at $\lambda=0.005$). The variant that reduces the hidden width to $16$ and enlarges the input embedding to $298$ trains for $14928.480\,\mathrm{s}$ ($0.298570\,\mathrm{s/epoch}$) and ends at a final loss of $0.0001570661$. The variant that keeps the original $32$-dimensional input embedding but appends an output head $16\!\to\!940\!\to\!2$ trains for $16181.268\,\mathrm{s}$ ($0.323625\,\mathrm{s/epoch}$) and ends at $0.00002591088$. The baseline LSTM-PINN records $19386.729\,\mathrm{s}$ ($0.387735\,\mathrm{s/epoch}$) and $0.0001005614$. We inspect the loss trajectories and the steady fields: the first variant exhibits abrupt regime changes with spikes in $10{,}000$–$20{,}000$ and $20{,}000$–$30{,}000$ steps; the second variant exhibits multiple spikes throughout training. These instabilities break the stable descent we require and do not yield robust steady-state structures. We therefore conclude that neither equal-capacity replacement qualifies as a baseline, and we retain the original LSTM-PINN as the baseline configuration.

\section{Ablation on Recurrent Backbones: RNN and Outer-Ring Residual vs. LSTM—Convergence Stability, Computational Cost, and Baseline Assessment}
We run an ablation under identical physics, sampling, optimizer, and schedule to isolate the effects of recurrent gating and an outer-ring residual. We fix the learning rate at $\lambda=5\times10^{-3}$, train for $50{,}000$ steps with Adam, keep the random seed and the interior/boundary batch sizes constant, and reuse the same spatial-to-sequence encoding used in the baseline. On $\Omega=[0,1]^2$ we predict $(u,v)$ from spatial input $(x,y)$, obtain derivatives by automatic differentiation, and form the momentum and incompressibility residuals:
\begin{equation}
    \begin{aligned}
r_x&=\rho_m\!\left(u\,\partial_x u+v\,\partial_y u\right)-\eta\left(\partial_{xx}u+\partial_{yy}u\right)-\tfrac{\eta}{3}\left(\partial_y u+\partial_x v\right)+\rho_e E_x,\\
r_y&=\rho_m\!\left(u\,\partial_x v+v\,\partial_y v\right)-\eta\left(\partial_{xx}v+\partial_{yy}v\right)-\tfrac{\eta}{3}\left(\partial_y u+\partial_y v\right)+\rho_e E_y,\\
r_c&=\partial_x u+\partial_y v.
\end{aligned}
\end{equation}
We enforce homogeneous Neumann conditions on all four edges:
\begin{equation}
    \partial_x u=\partial_x v=0\ \text{at}\ x\in\{0,1\},\qquad
\partial_y u=\partial_y v=0\ \text{at}\ y\in\{0,1\},
\end{equation}
and minimize the composite objective:
\begin{equation}
    \mathcal{L}_{\mathrm{total}}
=\mathrm{MSE}(r_x,0)+\mathrm{MSE}(r_y,0)+\mathrm{MSE}(r_c,0)+\mathcal{L}_{\mathrm{bnd}}.
\end{equation}
We resample interior and boundary points every step, log the linear/log-loss trajectories, wall-clock time and time-per-epoch, and render $u(x,y)$ and $v(x,y)$ on a high-resolution grid for side-by-side inspection.

We compare three recurrent designs while keeping the input embedding, hidden widths, and output head matched across models unless otherwise stated.

\textbf{(1) Ungated RNN.}
We remove gating and use a three-layer $\tanh$ RNN stack with hidden width $D$ to propagate the sequence features. At token $t$ and layer $\ell$ we update:
\begin{equation}
    h^{(\ell)}_{t}=\tanh\!\big(W^{(\ell)} x^{(\ell)}_{t}+U^{(\ell)} h^{(\ell)}_{t-1}+b^{(\ell)}\big),\quad
x^{(1)}_{t}=\phi(\xi_t),\quad x^{(\ell+1)}_{t}=h^{(\ell)}_{t},
\end{equation}
where $\phi(\cdot)$ denotes the input embedding of $(x,y)$ into $\mathbb{R}^{d_{\text{emb}}}$. A linear head maps $h^{(3)}_{t}$ to $[u_t,v_t]^\top$.

\textbf{(2) LSTM baseline.}
We use a three-layer LSTM with the same widths and embedding as the RNN setting. At token $t$ and layer $\ell$ the gates and states obey:
\begin{equation}
    \begin{aligned}
i_t^{(\ell)}&=\sigma(W_i^{(\ell)}x_t^{(\ell)}+U_i^{(\ell)}h_{t-1}^{(\ell)}+b_i^{(\ell)}),\quad
f_t^{(\ell)}=\sigma(W_f^{(\ell)}x_t^{(\ell)}+U_f^{(\ell)}h_{t-1}^{(\ell)}+b_f^{(\ell)}),\\
o_t^{(\ell)}&=\sigma(W_o^{(\ell)}x_t^{(\ell)}+U_o^{(\ell)}h_{t-1}^{(\ell)}+b_o^{(\ell)}),\quad
g_t^{(\ell)}=\tanh(W_c^{(\ell)}x_t^{(\ell)}+U_c^{(\ell)}h_{t-1}^{(\ell)}+b_c^{(\ell)}),
\end{aligned}
\end{equation}
\begin{equation}
    c_t^{(\ell)}=f_t^{(\ell)}\odot c_{t-1}^{(\ell)}+i_t^{(\ell)}\odot g_t^{(\ell)},\qquad
h_t^{(\ell)}=o_t^{(\ell)}\odot\tanh\!\big(c_t^{(\ell)}\big),
\end{equation}
with $x^{(1)}_{t}=\phi(\xi_t)$ and $x^{(\ell+1)}_{t}=h^{(\ell)}_{t}$. A linear head maps $h^{(3)}_{t}$ to $[u_t,v_t]^\top$.

\textbf{(3) LSTM with an outer-ring residual.}
We keep the LSTM baseline unchanged and inject the input embedding back to the last recurrent output through a parameter-free residual:
\begin{equation}
    y^{\mathrm{out}}_{t}\leftarrow y^{\mathrm{out}}_{t}+\alpha\,\Pi\!\big(\phi(\xi_t)\big),
\end{equation}
where $y^{\mathrm{out}}_{t}$ denotes the pre-head output of the last LSTM layer, $\Pi(\cdot)$ aligns dimensions by zero-padding or truncation when needed, and we set $\alpha=1$. This residual introduces no trainable parameters and tests whether re-injecting the embedded spatial signal at the output improves convergence or stability.

We keep the physics, sampling, optimizer, schedule, and logging identical across the three designs so that any difference in learning dynamics or steady-state fields arises from the recurrent unit (ungated vs. gated) and from the presence or absence of the outer-ring residual. We report loss histories (linear and logarithmic), wall-clock cost, time per epoch, final loss, and field visualizations for a controlled, reproducible comparison.
\begin{figure}[t]
    \centering
    \includegraphics[width=0.95\linewidth]{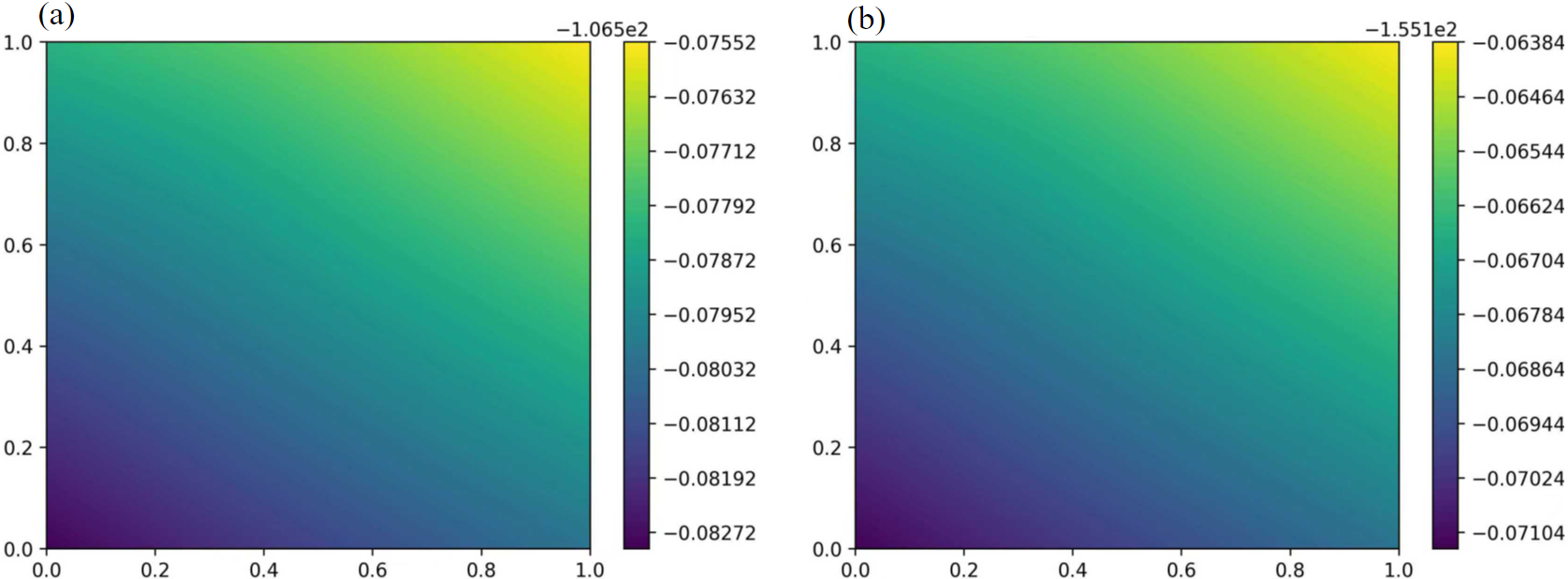}
    \caption{We render steady fields at step $50{,}000$ with the same setting: (a) $u(x,y)$ and (b) $v(x,y)$. Both maps exhibit smooth, diagonally oriented gradients over $\Omega=[0,1]^2$ without spurious boundary oscillations; the color scales show a narrow dynamic range with a sizable constant offset, so this configuration mainly captures near-linear spatial trends rather than rich structures.}
    \label{RNNR}
\end{figure}
\begin{figure}[t]
    \centering
    \includegraphics[width=0.95\linewidth]{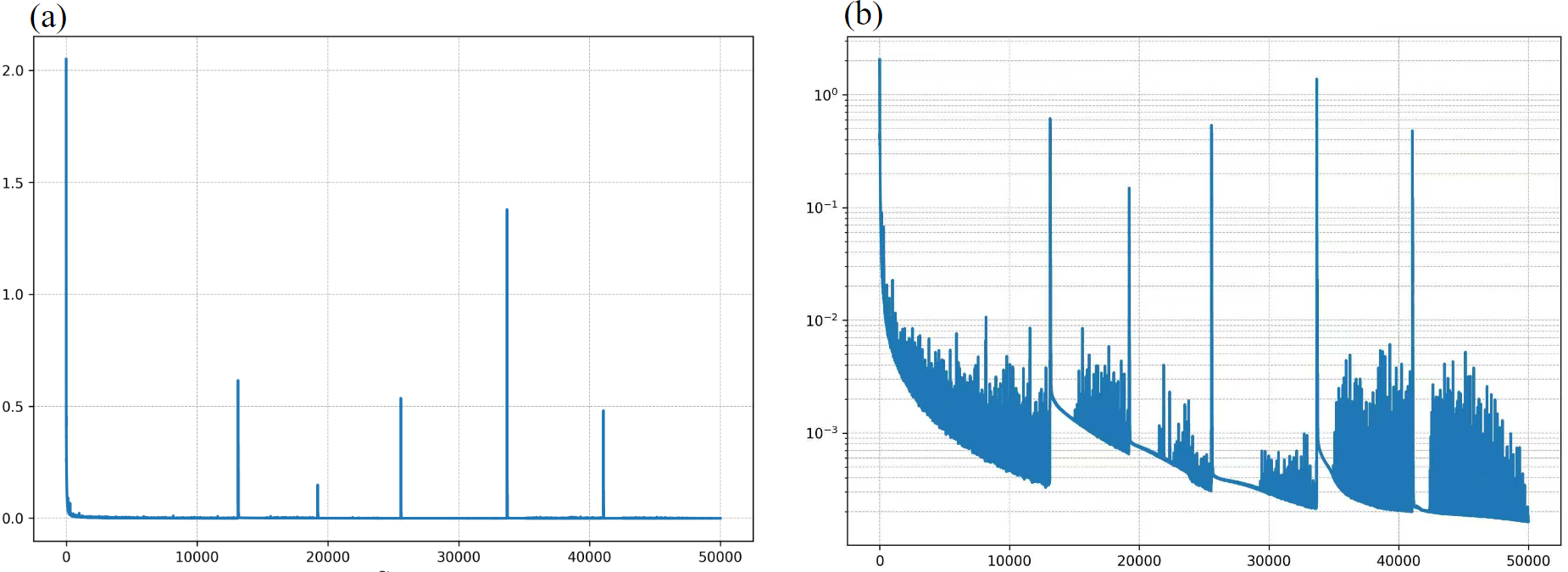}
    \caption{We embed $(x,y)$ and train a three-layer $\tanh$ RNN under the same physics losses, sampling, optimizer, and schedule ($\lambda=0.005$, $50{,}000$ steps). We plot the total loss versus step: (a) the linear axis shows a rapid drop from $\mathcal{O}(1)$ to the $10^{-2}$ range within the first few hundred steps, followed by bursty spikes near $1.2\times10^{4}$, $1.9\times10^{4}$, $2.6\times10^{4}$, $3.3\times10^{4}$, and $4.1\times10^{4}$; (b) the logarithmic axis shows piecewise descent punctuated by high-amplitude excursions and subsequent recoveries, which reveals instability of ungated recurrence over long spatial dependencies.}
    \label{RNNL}
\end{figure}
\begin{figure}[t]
    \centering
    \includegraphics[width=0.95\linewidth]{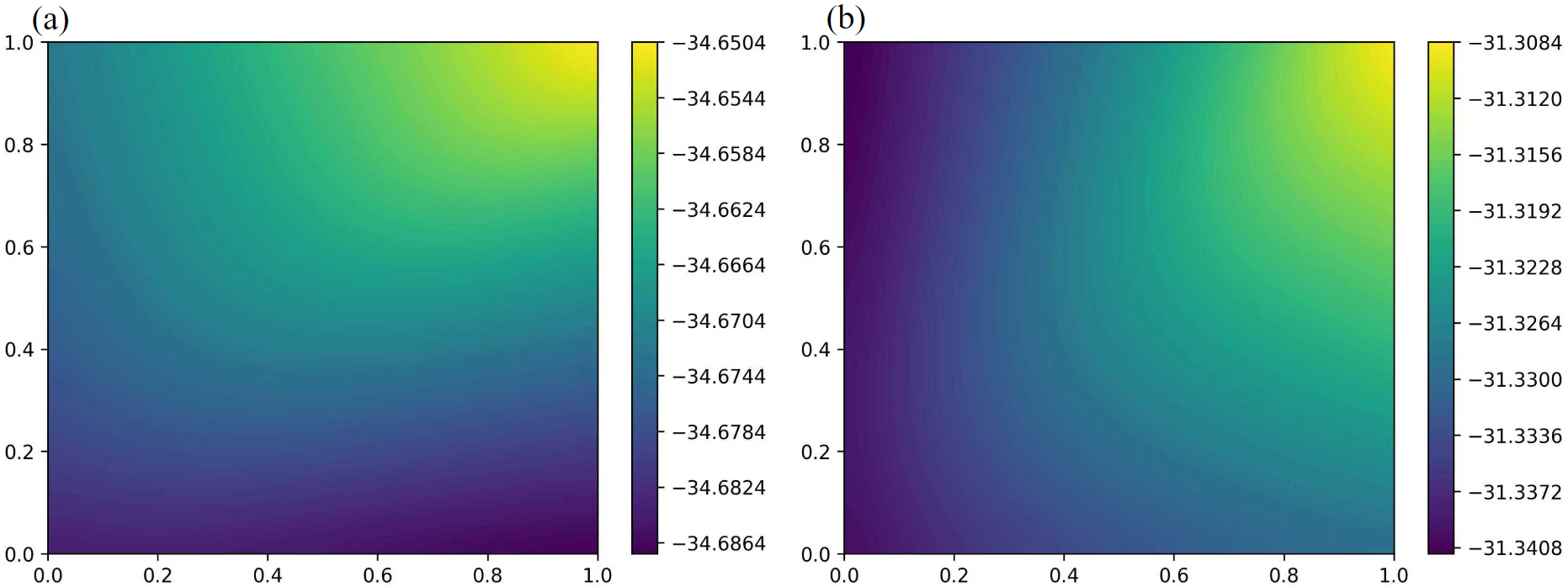}
    \caption{At step \(50{,}000\) we render the steady fields for the outer-ring residual model: (a) \(u(x,y)\) and (b) \(v(x,y)\). Both maps display smooth, diagonally oriented gradients over \(\Omega=[0,1]^2\) without spurious boundary oscillations; the color scales are narrow and carry a noticeable constant offset, indicating that this configuration mainly captures near-linear spatial trends rather than richer small-scale structures.}
    \label{RR}
\end{figure}
\begin{figure}[t]
    \centering
    \includegraphics[width=0.95\linewidth]{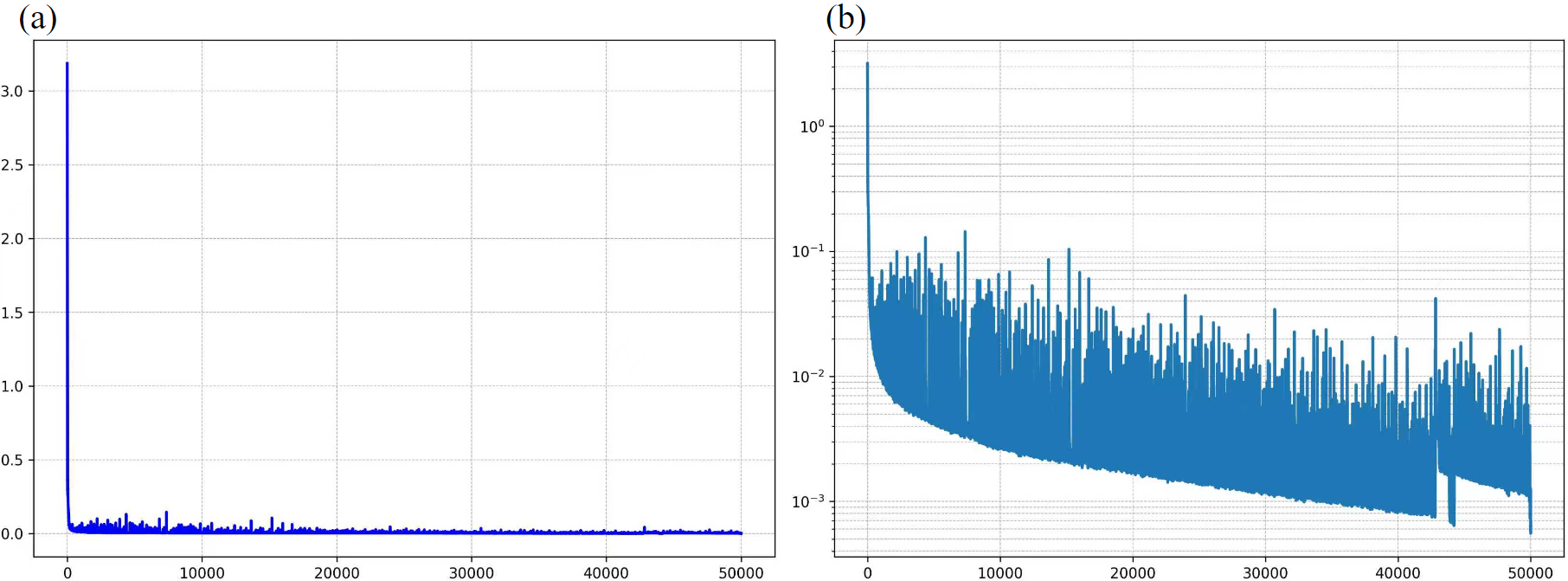}
    \caption{We evaluate the outer-ring residual LSTM-PINN under the same physics, sampling, optimizer, and schedule (\(\lambda=0.005\), \(50{,}000\) steps) and plot the total loss. (a) On the linear axis the loss drops from \(\mathcal{O}(1)\) to the \(10^{-2}\) range within the first few hundred steps, followed by several high-amplitude spikes and short plateaus; (b) on the logarithmic axis the curve exhibits piecewise monotone decay punctuated by large excursions, yielding a ``descent--spike--recovery'' rhythm and reaching the \(10^{-3}\) range near step \(50{,}000\). This pattern indicates that the outer-ring residual does not remove the instability induced by long-range dependencies.}
    \label{RL}
\end{figure}
\begin{table*}[t]
  \centering
  \caption{Comparison of recurrent alternatives at the fixed learning rate $\lambda=0.005$ with identical physics, sampling, optimizer, and hardware. Each model trains for $50{,}000$ epochs. We compare a three-layer $\tanh$ RNN, an outer-ring residual LSTM-PINN, and the baseline LSTM-PINN. We report the wall-clock training time (s), the mean time per epoch (s/epoch $=$ total$/50{,}000$), and the final loss (last item in the loss log). All numerics appear in decimal form (no scientific notation) to enable direct, reproducible comparison.}
  \label{tab:rnn_ring_lstm_fixedlr}
  \begin{tabular*}{\textwidth}{@{\extracolsep{\fill}} l r r r}
    \toprule
    Variant & training time (s) & time/epoch (s) & final loss \\
    \midrule
    RNN (three-layer $\tanh$) & 3066.079  & 0.061322 & 0.0001660037 \\
    LSTM-PINN with outer-ring residual & 17826.629 & 0.356533 & 0.0005623464 \\
    LSTM-PINN baseline & 19386.729 & 0.387735 & 0.0001005610 \\
    \bottomrule
  \end{tabular*}
\end{table*}

In Figures \ref{RNNR}-\ref{RL} and Table \ref{tab:rnn_ring_lstm_fixedlr}, under identical physics, sampling, and optimizer (\(\lambda=5\times10^{-3}\), \(50{,}000\) steps), we compare three recurrent backbones. The table shows: RNN — \(3066.079\,\mathrm{s}\), \(0.061322\,\mathrm{s/epoch}\), final loss \(0.0001660037\); LSTM with an outer-ring residual — \(17826.629\,\mathrm{s}\), \(0.356533\,\mathrm{s/epoch}\), final loss \(0.0005623464\); baseline LSTM-PINN — \(19386.729\,\mathrm{s}\), \(0.387735\,\mathrm{s/epoch}\), final loss \(0.0001005610\). The four plots confirm that the RNN exhibits too many abrupt spikes throughout training, while the outer-ring residual model shows frequent spikes in the 40k–50k window, disrupting late-stage convergence. In terms of cost, the outer-ring residual model is much more expensive than the RNN and still fails to improve the solution; compared with the baseline LSTM-PINN it attains a worse final loss and remains unstable. We therefore conclude that both alternatives are inferior to the LSTM-PINN baseline; subsequent studies should retain LSTM-PINN as the reference model.

\newpage

\section{Conclusion}

This study presents a systematic evaluation of Physics-Informed Neural Networks (PINNs) for modeling steady-state electrohydrodynamic flow in two-dimensional nanofluidic systems, comparing the performance of Long Short-Term Memory (LSTM) networks and Multi-Layer Perceptrons (MLPs). By embedding the coupled Navier-Stokes and electrostatic equations within the loss function, we demonstrate that while MLPs can achieve reasonable approximations under carefully tuned conditions, LSTM-based PINNs exhibit superior numerical stability, convergence robustness, and physical fidelity across a wider range of hyperparameters - advantages directly attributable to the LSTM's gating mechanisms that enable better gradient control and spatial information retention. These results establish that network architecture plays a pivotal role in solving PDE-constrained problems, particularly for multiscale systems with complex boundary conditions, where MLPs' limitations in handling stiff gradients become apparent while LSTMs provide a more robust and adaptable framework. We add two controlled ablations under identical physics and schedule ($\lambda=0.005$, $50{,}000$ steps). Equal-capacity swaps—hidden $=16$ with either a widened input embedding ($298$) or an enlarged output head ($16\!\to\!940\!\to\!2$)—trigger training spikes and never surpass the baseline. Backbone comparisons show that a three-layer ungated $\tanh$ RNN spikes repeatedly, while an outer-ring residual LSTM spikes densely in the $40$k–$50$k window and incurs higher cost. These results support retaining the LSTM-PINN as the reference model. Our findings highlight the significant potential of combining advanced neural architectures with physics-informed learning to develop mesh-independent computational alternatives, opening new possibilities for simulating transient, multiphase, and electrochemically active flows in complex device-scale applications.

\printcredits
\section*{Declaration of competing interest}
The authors declared that they have no conflicts of interest to this work. 

\section*{Acknowledgment}
This work is supported by the developing Project of Science and Technology of Jilin Province (20240402042GH). 

\section*{Data availability}
Data will be made available on request.

\clearpage
\bibliographystyle{cas-model2-names}
\bibliography{cas-refs}
\end{document}